\documentclass[reprint,amsmath,amssymb,aps,prb,superscriptaddress]{revtex4-2}
\usepackage{array}
\usepackage{bm}
\usepackage{xcolor}
\usepackage{float}
\usepackage{graphicx}
\usepackage{hyperref}
\hypersetup{colorlinks=true,allcolors=blue}
\usepackage{natbib}
\usepackage{newtxtext}
\usepackage{newtxmath}
\usepackage{mathrsfs}
\usepackage[caption=false]{subfig}

\begin{document}

\title{Topological phase diagrams of in-plane field polarized Kitaev magnets}

\author{Li Ern Chern}
\affiliation{T.C.M.~Group, Cavendish Laboratory, University of Cambridge, Cambridge CB3 0HE, United Kingdom}

\author{Claudio Castelnovo}
\affiliation{T.C.M.~Group, Cavendish Laboratory, University of Cambridge, Cambridge CB3 0HE, United Kingdom}


\begin{abstract}
While the existence of a magnetic field induced quantum spin liquid in Kitaev magnets remains under debate, its topological properties often extend to proximal phases where they can lead to unusual behaviors of both fundamental and applied interests. Subjecting a generic nearest neighbor spin model of Kitaev magnets to a sufficiently strong in-plane magnetic field, we study the resulting polarized phase and the associated magnon excitations. In contrast to the case of an out-of-plane magnetic field where the magnon band topology is enforced by symmetry, we find that it is possible for topologically trivial and nontrivial parameter regimes to coexist under in-plane magnetic fields. We map out the topological phase diagrams of the magnon bands, revealing a rich pattern of variation of the Chern number over the parameter space and the field angle. We further compute the magnon thermal Hall conductivity as a weighted summation of Berry curvatures, and discuss experimental implications of our results to planar thermal Hall effects in Kitaev magnets.
\end{abstract}


\maketitle

\begin{figure}
\subfloat[]{\label{figure:majoranaband}
\includegraphics[scale=0.18]{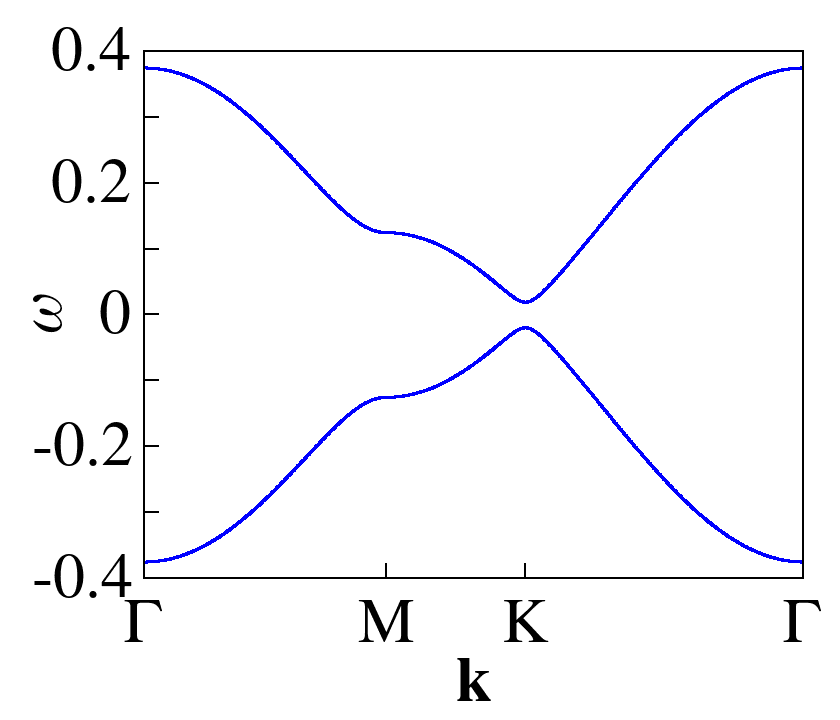}} \quad
\subfloat[]{\label{figure:fieldsphere}
\includegraphics[scale=0.3]{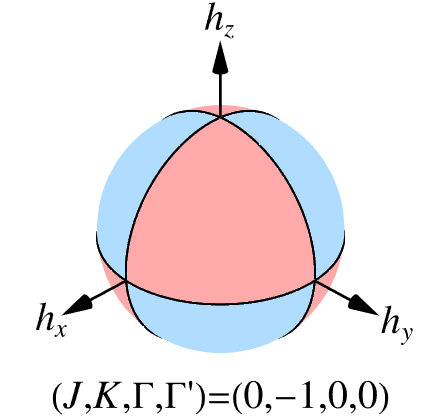}} \\
\subfloat[]{\label{figure:magnonband}
\includegraphics[scale=0.18]{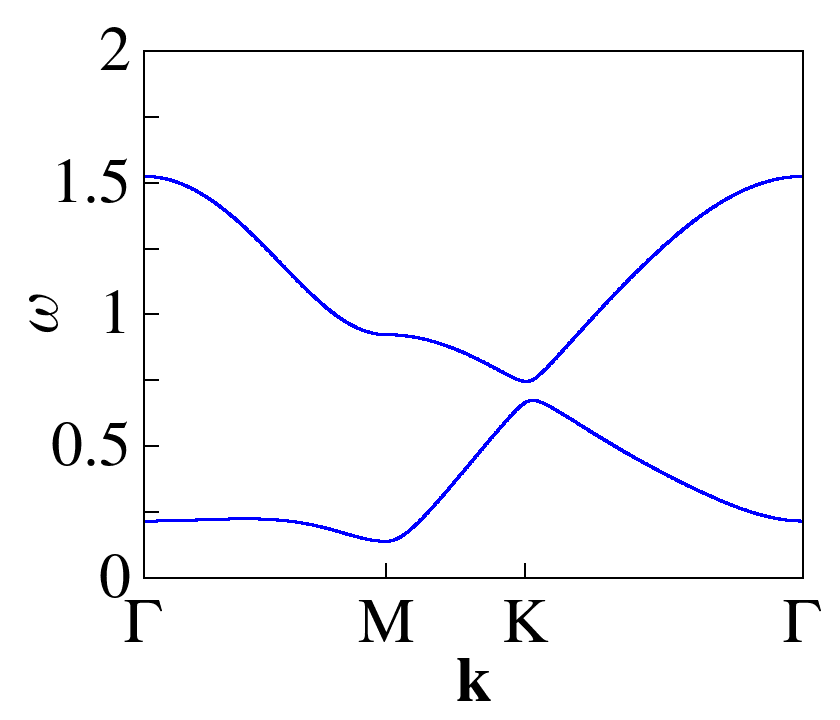}} \quad
\subfloat[]{\label{figure:couplingsphere}
\includegraphics[scale=0.3]{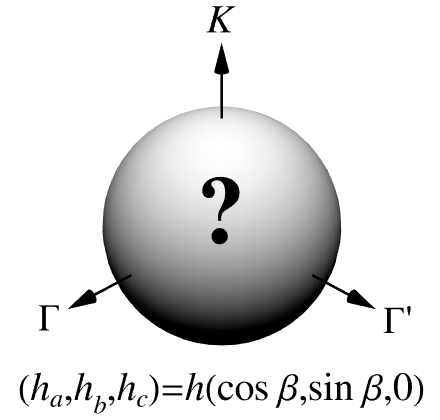}}
\caption{(a) Majorana spectrum of Kitaev honeycomb model in a perturbative magnetic field. (b) For the non-Abelian KSL, the Chern number of the lower Majorana band depends on the field direction through $\nu = \mathrm{sgn} (h_x h_y h_z)$. Red (blue) areas indicate $\nu = +1$ ($-1$), while black curves indicate the vanishing of the band gap. (c) Magnon spectrum of the polarized state in a realistic spin model \eqref{spinmodel} of Kitaev magnets under a magnetic field. (d) For the in-plane field polarized state, we find a nontrivial variation of the magnon Chern number over the parameter space and the field angle, see Figs.~\ref{figure:phasedeg000N}-\ref{figure:phasedeg300S}.}
\end{figure}

\textit{Introduction.}---Recently, there have been tremendous efforts in the search for Kitaev spin liquid (KSL) \cite{KITAEV20062} in candidate materials, ranging from iridates \cite{PhysRevB.82.064412,PhysRevLett.108.127203,nature25482} and ruthenium halides \cite{PhysRevB.90.041112,adma.202106831,PhysRevB.105.L041112} to cobaltates \cite{sciadv.aay6953,PhysRevB.102.224429}. These so-called Kitaev magnets \cite{Winter_2017,s42254-019-0038-2,TREBST20221,APLMaterials.10.080903} may realize a dominant Kitaev interaction $K$ via the Jackeli-Khaliullin \cite{PhysRevLett.102.017205} or related \cite{PhysRevB.97.014407,PhysRevB.97.014408,PhysRevLett.125.047201} mechanisms. Here, we focus on arguably the most popular amongst them, $\alpha$-RuCl$_3$ \cite{PhysRevResearch.2.033011}, whose zero-field ground state is a zigzag (ZZ) magnetic order \cite{PhysRevB.91.144420,PhysRevB.92.235119}, due to the presence of other symmetry-allowed interactions than $K$ \cite{PhysRevLett.112.077204}. However, an external magnetic field is found to promote a disordered phase, where a half-integer quantized thermal Hall conductivity $\kappa_{xy}^\mathrm{2D}/T = ( \nu / 2 ) ( \pi k_\mathrm{B}^2 / 6 \hbar ) $ \footnote{$\kappa_{xy}^\mathrm{2D} \equiv \kappa_{xy} d$ where $d$ is the interlayer distance between the honeycomb planes in a three dimensional structure.} is reported by Refs.~\cite{s41586-018-0274-0,PhysRevB.102.220404,science.aay5551,s41567-021-01501-y,PhysRevB.106.L060410}, hinting at chiral Majorana edge modes with Chern number $\nu=\pm 1$ in a non-Abelian KSL (Figs.~\ref{figure:majoranaband} and \ref{figure:fieldsphere}). It is also suggested that the field angle dependence of $\kappa_{xy}$ \cite{science.aay5551} or heat capacity \cite{s41467-021-27943-9,s41567-021-01488-6} can lend further support for the case of non-Abelian KSL. Meanwhile, other experiments \cite{s41567-021-01243-x,s41563-022-01397-w} report that $\kappa_{xy}$ in the field-induced phase behaves rather like a smooth function without any plateau, and decreases rapidly as the temperature approaches zero, which point to emergent heat carriers of bosonic nature.

While the existence of KSL at intermediate fields remains under debate \cite{s41467-019-08459-9,pnas.1821406116,s41467-019-10405-8,PhysRevB.100.165123,PhysRevLett.123.197201,PhysRevResearch.2.013014,s41467-020-15320-x,PhysRevResearch.2.043023}, Kitaev magnets eventually polarize at sufficiently high fields, where the collective excitations are magnons, which can give rise to experimentally measurable transport signals. Furthermore, if the magnon bands are topological, the resulting thermal Hall conductivity can reach the same order of magnitude as the half quantized value \cite{PhysRevLett.126.147201,PhysRevB.103.174402,s41535-021-00331-8,PhysRevB.107.184418}. Although for magnons $\kappa_{xy}$ at low temperatures is not directly proportional to the Chern number $\nu$ of the lower band, the latter is very often a good indicator of the opposite sign of the former. Therefore, phase diagrams that reveal the magnon Chern number across generic model parameters of Kitaev magnets \cite{1807.05232} are valuable to identify topological magnons and to interpret thermal transport measurements at high fields (Figs.~\ref{figure:magnonband} and \ref{figure:couplingsphere}). The main objective of this Letter is precisely to present such topological phase diagrams for in-plane magnetic fields, which are relevant to experiments of planar thermal Hall effect \cite{science.aay5551,s41567-021-01501-y,s41567-021-01243-x,s41563-022-01397-w,PhysRevResearch.4.L042035}.

We note that Kitaev magnets such as $\alpha$-RuCl$_3$ are polarized more easily by in-plane fields than out-of-plane fields, likely due to an anisotropic $g$ tensor \cite{PhysRevB.91.094422,PhysRevB.94.064435,srep37925,PhysRevLett.120.077203} and a positive $\Gamma$ interaction \cite{s41567-020-0874-0}, which discounts the out-of-plane field strength and disfavors an out-of-plane magnetization, respectively \cite{PhysRevB.103.174402,chernthesis}. The case of polarizing Kitaev magnets with strong out-of-plane fields has been studied theoretically in Ref.~\cite{PhysRevB.98.060404} (see also Ref.~\cite{PhysRevB.98.060405}). It is found that, within the linear spin wave approximation, the $J K \Gamma \Gamma'$ model can be effectively reduced to a $JK$ model. The $C_3$ symmetry also plays an important role in the diagnosis of magnon band topology in Kitaev magnets, based on topological quantum chemistry or symmetry indicator theory \cite{s41467-017-00133-2,nature23268,PhysRevX.7.041069,sciadv.aat8685,s41467-021-26241-8}. As demonstrated in Ref.~\cite{PhysRevLett.130.206702}, the magnon bands must be topological whenever a gap exists in between.

In this Letter, we consider the nearest neighbor $J K \Gamma \Gamma'$ model polarized by in-plane magnetic fields, which break the $C_3$ symmetry, and map out the phase diagrams of topological magnons. Unlike the aforementioned case, none of the model parameters can be made redundant. We find that, as long as the field is not along the armchair direction, there exist parameter regions that are topological ($\nu = \pm 1$) as well as trivial ($ \nu = 0 $) ones, the latter of which can be understood via an effective Hamiltonian \cite{PhysRevB.98.060404}. We discuss the implications of our results to thermal Hall conductivities of Kitaev magnets at high fields, from which we propose a scheme to determine the relevant candidate parametrizations.

\begin{figure}
\subfloat[]{\label{figure:geometry}
\includegraphics[scale=0.34]{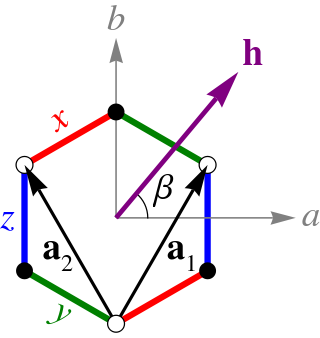}} \qquad
\subfloat[]{\label{figure:symmetry}
\includegraphics[scale=0.23]{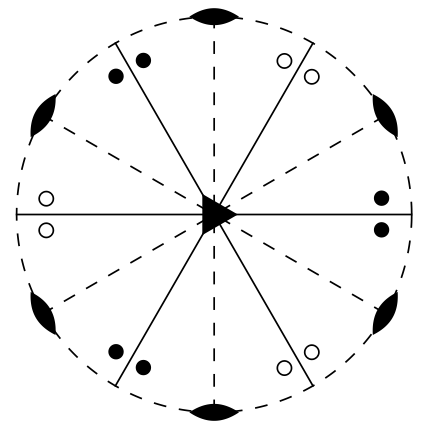}}
\caption{(a) The three bond types $x$, $y$, and $z$ in Kitaev magnets, the in-plane crystallographic axes $a$ and $b$, and the primitive lattice vectors $\mathbf{a}_1$ and $\mathbf{a}_2$. An external magnetic field $\mathbf{h}$ is applied in-plane at the azimuthal angle $\beta$. (b) The $\bar{3}m$ point group of the $J K \Gamma \Gamma'$ model. If $\mathbf{h}$ transforms under a symmetry element that maps a filled circle to an empty circle or vice versa, then the Chern number $\nu$ flips sign. If one circle is mapped to another of the same type, then $\nu$ remains invariant.}
\end{figure}

\textit{Model.}---The most generic nearest neighbor spin Hamiltonian for Kitaev magnets is the $J K \Gamma \Gamma'$ model \cite{PhysRevLett.112.077204}. In an external magnetic field $\mathbf{h}$, it reads
\begin{equation} \label{spinmodel}
\begin{aligned}[b]
H = & \sum_{\lambda = x,y,z} \sum_{\langle ij \rangle \in \lambda} \big[ J \mathbf{S}_i \cdot \mathbf{S}_j + K S_i^\lambda S_j^\lambda + \Gamma ( S_i^\mu S_j^\nu + S_i^\nu S_j^\mu ) \\
& + \Gamma' (S_i^\mu S_j^\lambda + S_i^\lambda S_j^\mu + S_i^\nu S_j^\lambda + S_i^\lambda S_j^\nu) \big] - \sum_i \mathbf{h} \cdot \mathbf{S}_i ,
\end{aligned}
\end{equation}
where $(\lambda, \mu, \nu)$ is a cyclic permutation of $(x, y, z)$. For convenience of analysis, we write the field strength $\lvert \mathbf{h} \rvert \equiv h S$ in terms of the spin magnitude $S \equiv \lvert \mathbf{S}_i \rvert$ \footnote{We also take $S$ to be dimensionless so that $J, K, \Gamma, \Gamma'$ and $h$ have units of energy.}. An in-plane field can be parametrized as $( h_a , h_b, h_c ) = h (\cos \beta, \sin \beta, 0)$, where $\beta \in [0, 2 \pi)$ is the azimuthal angle in the honeycomb plane, see Fig.~\ref{figure:geometry} \footnote{The crystallographic $abc$ axes, which distinguish the in-plane and out-of-plane directions, and the cubic $xyz$ axes, according to which the spin components in \eqref{spinmodel} are defined, are related by $a \parallel [11\bar{2}]$, $b \parallel [\bar{1}10]$, and $c \parallel [111]$.}. We apply linear spin wave theory \cite{PhysRev.58.1098,Jones_1987} to the in-plane field polarized state of \eqref{spinmodel}, and obtain an analytical expression for the magnon spectrum $\omega_\pm (\mathbf{k}) = S \sqrt{E(\mathbf{k}) \pm \Delta (\mathbf{k})} / 2 $, where
\begin{widetext}
\begin{subequations}
\begin{align}
& E(\mathbf{k}) = 4 (h-c_1)^2 + \big\lvert c_2 f_\mathbf{k} + c_4 g_\mathbf{k} \big\rvert^2 - \big\lvert c_5 f_\mathbf{k} - c_4 g_\mathbf{k} \big\rvert^2 - 4 \big\lvert c_6 (f_\mathbf{k} - 3) + c_7 g_\mathbf{k} \big\rvert^2 + 2 c_3 (c_2 + c_5) \mathrm{Re} \big[ f_\mathbf{k} \big] , \label{Ekdefine} \\
\begin{split} \label{Dksquaredefine}
& \Delta^2 (\mathbf{k}) = 16 (h - c_1)^2 \big\lvert c_2 f_\mathbf{k} + c_3 + c_4 g_\mathbf{k} \big\rvert^2 - 4 (c_2 + c_5)^2 \big\{ \mathrm{Im} \big[ f_\mathbf{k} (c_3 + c_4 g_\mathbf{k}^*) \big] \big\}^2 - 16 (c_2^2 - c_5^2) \big\{ \mathrm{Im} \big[ f_\mathbf{k} (3 c_6 - c_7 g_\mathbf{k}^*) \big] \big\}^2 \\
& \qquad \quad \; \: - 32 (c_2 + c_5) \mathrm{Im} \big[ f_\mathbf{k} (3 c_6 - c_7 g_\mathbf{k}^*) \big] \mathrm{Im} \big[ c_6 f_\mathbf{k} (c_3 + c_4 g_\mathbf{k}^*) + g_\mathbf{k} (c_3 c_7 + 3 c_4 c_6) \big] , 
\end{split} \\
\begin{split} \label{cidefine}
& c_1 = 3 J + K - \Gamma - 2 \Gamma' , 
c_2 = \frac{1}{6} \left[ 12 J + 4 K + 2 \Gamma + 4 \Gamma' + (K + 2 \Gamma - 2 \Gamma') \cos (2 \beta) \right] , c_3 = - \frac{\cos (2 \beta) }{2} \left( K + 2 \Gamma - 2 \Gamma' \right) , \\
& c_4 = \frac{\sin (2 \beta)}{2 \sqrt{3}} \left( K + 2 \Gamma - 2 \Gamma' \right) , 
c_5 = \Gamma + 2 \Gamma' - \frac{\cos (2 \beta)}{6} \left( K + 2 \Gamma - 2 \Gamma' \right) , c_6 = \frac{\sin \beta}{3 \sqrt{2}} \left( K - \Gamma + \Gamma' \right) , 
c_7 = \frac{\cos \beta}{\sqrt{6}} \left( K - \Gamma + \Gamma' \right) ,
\end{split}
\end{align}
\end{subequations}
\end{widetext}
$ f_\mathbf{k} = 1 + \exp ( i k_1 ) + \exp ( i k_2 ) $, $ g_\mathbf{k} = \exp ( i k_1 ) - \exp( i k_2 ) $, and $ k_1, k_2 \in [ 0, 2 \pi ) $ are components of the crystal momentum defined according to $\mathbf{a}_1 , \mathbf{a}_2$ in Fig.~\ref{figure:geometry}. Let $\Delta (\mathbf{k}) = \sqrt{\Delta^2 (\mathbf{k})} \geq 0$, so that $\omega_- (\mathbf{k})$ [$\omega_+ (\mathbf{k})$] corresponds to the lower (upper) band. For clarity, we refer to the gap between the two bands, $\min_\mathbf{k} [\omega_+ (\mathbf{k}) - \omega_- (\mathbf{k})] \geq 0$, as the \textit{band gap}, which is not to be confused with the \textit{excitation gap}, $\min_\mathbf{k} \omega_- (\mathbf{k}) > 0$. Chern number is a topological invariant that can never change as long as a finite band gap is maintained \cite{annurev-conmatphys-031620-104715}, i.e., a topological phase transition can only occur when $\Delta (\mathbf{k}) = 0$ for some $\mathbf{k}$.

We assume a polarized state in which the excitation gap grows with $h$, so that the system becomes more stable as $h$ increases, rather than undergoing a magnon instability. This requires $h>c_1$ \cite{supplement}, from which we deduce the following. For a given set of parameters $ \lbrace J, K,\Gamma, \Gamma' \rbrace $, if the band gap is finite (zero), then it remains finite (zero) as $h$ varies, unless $h \longrightarrow \infty$. Therefore, the topological phase diagrams are independent of the field strength, and, for a given field angle, we can map them out by first solving for the zeros of \eqref{Dksquaredefine} and then choosing a sufficiently high field to compute the Chern numbers \cite{bernevigtextbook,vanderbilttextbook,JPSJ.74.1674,s12200-019-0963-9} at parameters away from these zeros.

\textit{Topological phase diagrams.}---For finite in-plane fields, the band gap closes iff the set of parameters $ \lbrace J, K,\Gamma, \Gamma' \rbrace $ meets any of the criteria listed in Table \ref{table:gapless}. Whenever the band gap is finite, let the Chern number of the lower (upper) band be $\nu$ ($-\nu$), which transforms according to the $A_{2g}$ representation of the point group $\bar{3}m$ \cite{bradleytextbook,dresselhaustextbook}, and flips sign under time reversal \cite{PhysRevB.103.174402}, as in the case of the non-Abelian KSL \cite{s41467-021-27943-9}. More specifically, fixing the couplings, (i) $\nu \longrightarrow \nu$ if $\mathbf{h}$ is rotated by $2 \pi /3$ about the out-of-plane axis, (ii) $\nu \longrightarrow -\nu$ if $\mathbf{h}$ is rotated by $\pi$ about the $b$ axis, and (iii) $\nu \longrightarrow -\nu$ if $\mathbf{h} \longrightarrow - \mathbf{h}$, while the phase boundaries are invariant under these actions \cite{supplement}. Hence, $\beta \in [0 , \pi / 6]$ serves as an independent unit, to which all other angles can be related by symmetries, see Fig.~\ref{figure:symmetry}. On the other hand, flipping the signs of all couplings leaves $\nu$ invariant \cite{supplement}.

\begin{table}
\caption{\label{table:gapless}For field angles $ 0 \leq \beta < \pi / 6 $, the band gap closes iff the parameters of the $J K \Gamma \Gamma'$ model satisfy any of the following equations. For $\beta = \pi / 6$, the band gap is zero whenever (I) or \eqref{solubility} is satisfied.}
\begin{ruledtabular}
\begin{tabular}{cc}
I & $K + 2 \Gamma - 2 \Gamma' = 0$ \\
II & $6 J + 2 K + \Gamma + 2 \Gamma' = 0$ \\
III & $6 J + 2 K + \Gamma + 2 \Gamma' + 2 ( K + 2 \Gamma - 2 \Gamma' ) \cos (2 \beta) = 0$ \\
IV & $6 J + 2 K + \Gamma + 2 \Gamma' - 2 ( K + 2 \Gamma - 2 \Gamma' ) \cos (2 \beta + \pi / 3) = 0$ \\
V & $6 J + 2 K + \Gamma + 2 \Gamma' - 2 ( K + 2 \Gamma - 2 \Gamma' ) \cos (2 \beta - \pi / 3) = 0$ \\
VI & $3 J + K + 2 \Gamma + 4 \Gamma' = 0$ \; if \eqref{solubility} holds \\
VII & $K - \Gamma + \Gamma' = 0$ \; if \eqref{solubility} holds
\end{tabular}
\end{ruledtabular}
\end{table}

\begin{figure}
\subfloat[]{\label{figure:phasedeg000N}
\includegraphics[scale=0.35]{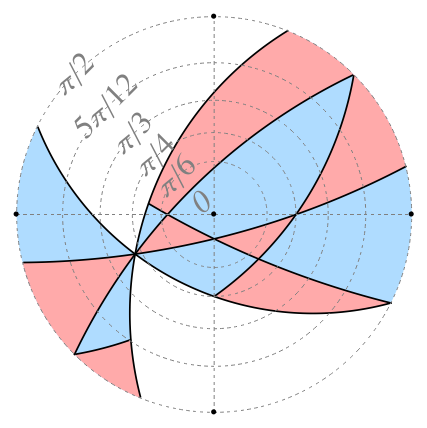}} \quad
\subfloat[]{\label{figure:phasedeg000S}
\includegraphics[scale=0.35]{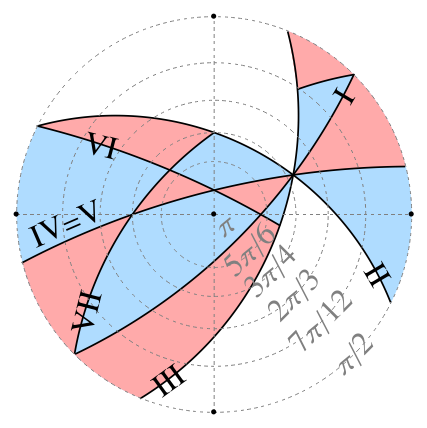}} \\
\subfloat[]{\label{figure:phasedeg075S}
\includegraphics[scale=0.35]{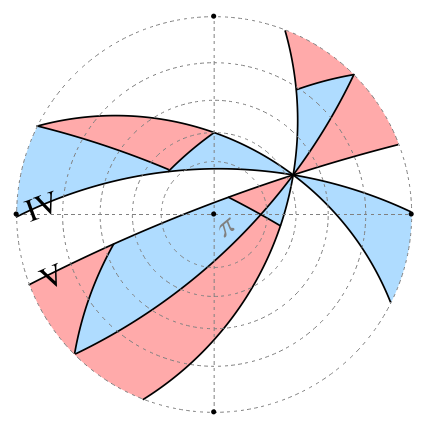}} \quad
\subfloat[]{\label{figure:phasedeg150S}
\includegraphics[scale=0.35]{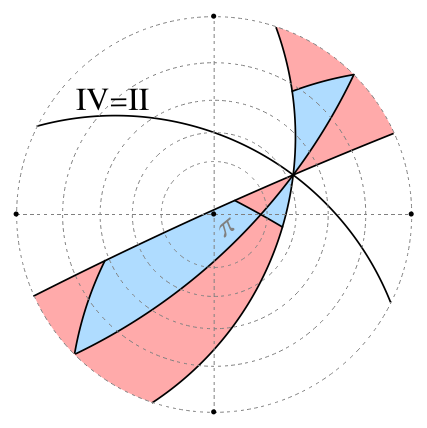}} \\
\subfloat[]{\label{figure:phasedeg225S}
\includegraphics[scale=0.35]{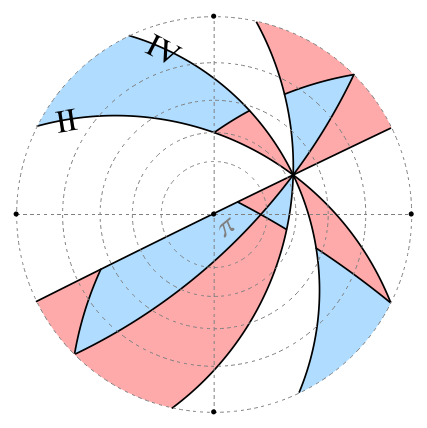}} \quad
\subfloat[]{\label{figure:phasedeg300S}
\includegraphics[scale=0.35]{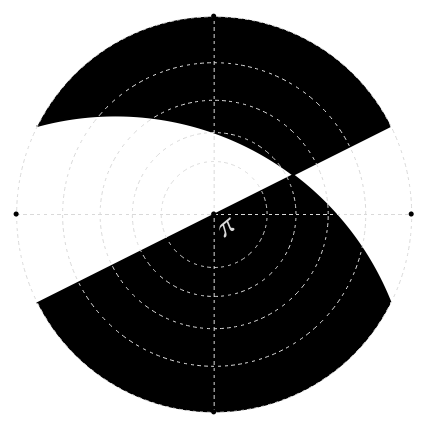}}
\caption{Topological phase diagrams of the in-plane field polarized states at $\beta$ equal to (a,b) $0$, (c) $\pi / 24$, (d) $\pi / 12$, (e) $\pi / 8$, and (f) $\pi / 6$, over the space of couplings parametrized by $( J , K , \Gamma , \Gamma' ) = ( 0 , \cos \theta , \sin \theta \cos \phi , \sin \theta \sin \phi )$. Red, white, and blue areas indicate the Chern number of the lower magnon band $\nu = +1$, $0$, and $-1$, respectively, while black curves or areas indicate the vanishing of the band gap. Roman numerals label the phase boundaries as in Table \ref{table:gapless}. Gray dashed circles indicate constant latitudes $\theta$. In each diagram, the center is the $K=\pm 1$ ($\theta = 0$ or $\pi$) limit, while the left/right and top/bottom ends on the equator ($\theta = \pi / 2$) are the $ \Gamma = \pm 1 $ ($\phi = 0$ or $\pi$) and $ \Gamma' = \pm 1 $ ($ \phi = \pi / 2 $ or $3 \pi / 2$) limits, respectively.}
\end{figure}

For visualizations, we set $J=0$ and calculate $\nu$ over the spherical parameter space defined by $K^2 + \Gamma^2 + {\Gamma'}^2 = 1$, at the field angles $\beta=0, \pi / 24, \pi / 12, \pi / 8, \pi / 6$, see Figs.~\ref{figure:phasedeg000N}-\ref{figure:phasedeg300S} \footnote{We have used the stereographic projection $( \Gamma / ( 1 \pm K ), \Gamma' / ( 1 \pm K) )$, where the plus (minus) sign is for the $K>0$ ($K<0$) hemisphere.}. We make two observations, with the understanding that all angles mentioned below are defined modulo $\pi / 3$. First, for $\beta \neq \pi / 6$, there exist both parameter regions with topological magnons and those without. For $\beta = \pi / 6$, topological magnons are altogether forbidden due to a $C_2$ symmetry \cite{PhysRevLett.126.147201,PhysRevResearch.3.013179}. Second, the total area $A$ of the parameter regions with $\nu=\pm 1$ is maximal at $\beta = 0$, which implies that, for a Kitaev magnet dominated by nearest neighbor anisotropic interactions, topological magnons are most likely found when the in-plane field is along the $a$ axis \footnote{A plot of $A ( \beta ) $ is shown in the Supplemental Material \cite{supplement}.}.

To understand why magnons are topologically trivial in certain parameter regions, we analyze the linear spin wave theory at high fields by systematically integrating out the pairing terms \cite{PhysRevB.98.060404}. This is achieved via a Schrieffer-Wolff transformation \cite{BRAVYI20112793}, from which we obtain an effective hopping model of the form $ \mathcal{H}^\mathrm{eff} ( \mathbf{k} ) = d_0 ( \mathbf{k} ) \mathbf{1}_{2 \times 2} + \mathbf{d} ( \mathbf{k} ) \cdot \bm{\sigma} $. The band gap vanishes iff $\mathbf{d} ( \mathbf{k} ) = \mathbf{0}$. When $ \mathbf{d} ( \mathbf{k} ) \neq \mathbf{0} $, the Chern number of the lower band is given by the winding number of the map $ \hat{\mathbf{d}} (\mathbf{k}) \equiv \mathbf{d} ( \mathbf{k} ) / \lvert \mathbf{d} ( \mathbf{k} ) \rvert$ from the Brillouin zone to a sphere \cite{bernevigtextbook},
\begin{equation} \label{winding}
\nu = \frac{1}{4 \pi} \int_\mathrm{FBZ} \mathrm{d}^2 \mathbf{k} \left[ \hat{\mathbf{d}} (\mathbf{k}) \cdot \frac{ \partial \hat{\mathbf{d}} ( \mathbf{k} ) } { \partial k_x } \times \frac{ \partial \hat{\mathbf{d}} ( \mathbf{k} ) } { \partial k_y } \right] .
\end{equation}
One finds that the third component of $\mathbf{d} ( \mathbf{k} )$ vanishes throughout the Brillouin zone when $K - \Gamma + \Gamma' = 0$ \cite{supplement}, which defines the phase boundary (VII) within the parameter region
\begin{equation} \label{solubility}
\big\lvert \lvert c_2 + c_4 \rvert - \lvert c_2 - c_4 \rvert \big\rvert \leq \lvert c_2 + c_3 \rvert \leq \lvert c_2 + c_4 \rvert + \lvert c_2 - c_4 \lvert .
\end{equation}
On the other hand, there exist parameters outside \eqref{solubility} that satisfy $K - \Gamma + \Gamma' = 0$ and possess a finite band gap simultaneously \footnote{One can imagine the extension of (VII) to the white areas in Figs.~\ref{figure:phasedeg000N}-\ref{figure:phasedeg225S}.}. At these parameters, the triple product in \eqref{winding} is identically zero, and consequently $\nu=0$. Any other parameter that can be continuously connected to these parameters without a gap closing must be topologically trivial as well.

\begin{figure}
\subfloat[]{\label{figure:parameter}
\includegraphics[scale=0.34]{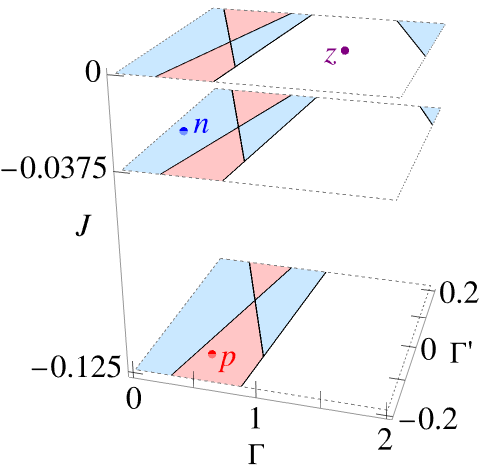}} \quad
\subfloat[]{\label{figure:hall}
\includegraphics[scale=0.3]{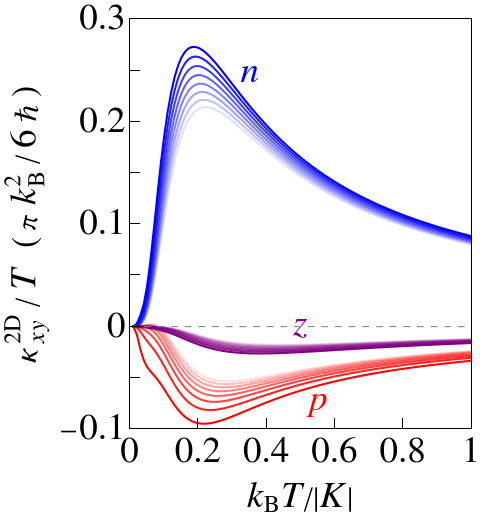}}
\caption{(a) Candidate parametrizations $p$ \cite{PhysRevB.93.155143}, $n$ \cite{PhysRevB.97.134424}, and $z$ \cite{PhysRevLett.118.107203} of $\alpha$-RuCl$_3$, with $K$ set to $-1$ and other interactions scaled accordingly, and topological phase diagrams in their neighborhoods under a magnetic field $\mathbf{h} \parallel a$. (b) Thermal Hall conductivities of $p$, $n$, and $z$ due to magnons in the polarized state, at field strengths $h$ starting from $0.18$, $0.10$, and $0.89$, respectively, and increasing to $0.32$, $0.24$, and $1.03$ in steps of $0.02$. Lighter colors indicate higher fields. $S=1/2$ is used.}
\end{figure}

\textit{Thermal Hall effect.}---We discuss how the topological phase diagrams relate to experimentally measurable quantities by connecting the Chern number to the thermal Hall conductivity \footnote{We assume that the heat current is applied along the $a$ axis, while the transverse temperature gradient is measured along the $b$ axis, as in the experiments.}, which is given by \cite{PhysRevLett.106.197202,PhysRevB.89.054420,JPSJ.86.011010}
\begin{equation} \label{thermalhallconductivity}
\kappa_{xy} = - \frac{k_\mathrm{B}^2 T}{\hbar V} \sum_{n=1}^\mathcal{N} \sum_{\mathbf{k} \in \mathrm{FBZ}} c_2 \left[ g \left( \frac{ \hbar \omega_{n \mathbf{k}}}{k_\mathrm{B} T} \right) \right] \Omega_{n \mathbf{k}} 
\end{equation}
for magnons, where $n$ is the band index ranging from $1$ to $\mathcal{N}=2$, $ c_2 (x) = \int_0^x \mathrm{d} t \, \ln^2 [ ( 1 + t ) / t ] $, $g (x) = 1 / (e^x - 1)$, and $\Omega_{n \mathbf{k}}$ is the momentum space Berry curvature \cite{supplement}. While the Chern number $\nu_n$ is given by the summation of $\Omega_{n \mathbf{k}}$ over $\mathbf{k}$, $\kappa_{xy}$ is given by a weighted summation of $\Omega_{n \mathbf{k}}$ with \textit{non-positive} weights. Also, high-energy magnons contribute less to $\kappa_{xy}$ than low-energy ones. Therefore, though $\kappa_{xy}$ is not directly proportional to $\nu$, one can very often use the latter to infer the sign of the former at low temperatures. More precisely, $\nu > 0$ ($ \nu < 0$) means that there is an excess of positive (negative) Berry curvatures in the lower band, and by \eqref{thermalhallconductivity} the sign of $\kappa_{xy}$ is expected to be opposite to $\nu$ \footnote{There may be a situation where $\nu > 0$ but $\Omega_{n \mathbf{k}} < 0$ is concentrated at the lowest energy magnons, so that $\kappa_{xy}$ have the same sign as $\nu$ in a small temperature window near $T=0$. However, if we examine an extended temperature range, we have predominantly $\mathrm{sgn} (\kappa_{xy}) = - \mathrm{sgn} (\nu)$.}. On the other hand, $\nu = 0$ means that the net Berry curvature is zero, so $\kappa_{xy}$ is generically small though not necessarily zero, and its sign is arbitrary.

We illustrate these ideas with three proposed parametrizations of $\alpha$-RuCl$_3$ in the literature, $( J , K , \Gamma , \Gamma' ) = (-1,-8,4,-1) $ \cite{PhysRevB.93.155143}, $(-1.5, -40, 5.3, -0.9)$ \cite{PhysRevB.97.134424}, and $(0, -6.8 , 9.5 , 0)$ \cite{PhysRevLett.118.107203}, where energies are given in units of meV. For $\mathbf{h} \parallel a$, these parametrizations are located in the $\nu=+1$, $-1$, and $0$ regimes, respectively, so we label them by $p$, $n$, and $z$, see Fig.~\ref{figure:parameter}. For each of them, we calculate $\kappa_{xy}$ as a function of $T$ at several values of $h$, see Fig.~\ref{figure:hall}. We find that $\kappa_{xy}$ is negative (positive) for $p$ ($n$) as expected, while $\kappa_{xy}$ for $z$ is several times smaller. If we assume that the measured $\kappa_{xy} > 0$ in the field-induced phase under $\mathbf{h} \parallel - a$ in $\alpha$-RuCl$_3$ \cite{science.aay5551} is indeed determined by a dominant magnon contribution, then $p$ appears to be a more promising candidate parametrization. We also list three criteria that are conducive for a large magnon thermal Hall effect, which help us to understand the difference in $\kappa_{xy}$ between the three parametrizations, as follows. (i) The bands are topological. (ii) The excitation gap is not too large, so that the lower band is thermally populated at low temperatures. (iii) The band gap is not too small, so that the population of the upper band remains negligible over an extended temperature range. For instance, at the respective lowest fields, the excitation gaps of $p$, $n$, and $z$ are $0.16$, $0.19$, and $0.24$, while the band gaps are $0.07$, $0.25$, and $0.78$, in units of $\lvert K \rvert S$. $n$ and $p$ fulfill (i) and are comparable in (ii), but $n$ does better than $p$ in (iii), so $n$ yields a larger $\kappa_{xy}$. On the other hand, $z$ is comparable to $p$ and $n$ in (ii) and does better in (iii), but $z$ fails (i), so its $\kappa_{xy}$ is small. As $h$ increases, the excitation gap becomes larger and $\kappa_{xy}$ decreases.

\textit{Discussion.}---In summary, we have mapped out topological phase diagrams of Kitaev magnets polarized by in-plane magnetic fields, which reveal the magnon Chern number over a large parameter space. Since topological magnons are generally expected to yield a sizable thermal Hall conductivity with sign opposite to the Chern number at low temperatures, our results will be helpful in determining the relevant parametrizations of Kitaev magnets including $\alpha$-RuCl$_3$. We briefly address the effects of the third nearest neighbor Heisenberg exchange \cite{PhysRevB.93.214431,s41467-017-01177-0} and the magnon interactions \cite{PhysRevB.79.144416,RevModPhys.85.219,PhysRevB.88.094407,PhysRevB.93.014418,PhysRevB.102.155134,PhysRevB.106.094431,PhysRevB.106.214411,PhysRevB.109.014424} in the Supplemental Material \cite{supplement}. While the window of a field-induced KSL might be shut in many of the candidate materials, the door to topological magnons is most likely open and accessible via high fields. We appreciate that alternative sources of heat carriers in Kitaev magnets, such as spinons \cite{PhysRevResearch.1.013014,PhysRevResearch.2.033283,PhysRevB.104.195149}, triplons \cite{PhysRevLett.122.177201}, phonons \cite{PhysRevX.12.021025}, and visons \cite{2206.14197}, as well as some effects arising from spin-lattice coupling \cite{PhysRevX.8.031032,PhysRevLett.121.147201,PhysRevResearch.2.033180,s41467-021-23826-1,PhysRevB.106.024413,2301.07401}, have been proposed. One particularly interesting future direction is to investigate the interplay between different types of topological excitations, whether they cooperate with one another and lead to a large thermal Hall conductivity \cite{PhysRevB.108.L140402} or other unusual properties.

We thank Kyusung Hwang, Hana Schiff, and Robert-Jan Slager for useful discussions. This work was supported by Engineering and Physical Sciences Research Council Grants No.~EP/T028580/1 and No.~EP/V062654/1.

\bibliography{reference231015}

\clearpage

\onecolumngrid

\begin{center}
\textbf{\large Supplemental Material: Topological phase diagrams of in-plane field polarized Kitaev magnets}
\end{center}
\begin{center}
Li Ern Chern$^1$ and Claudio Castelnovo$^1$
\end{center}
\begin{center}
{\small
\textit{$^1$T.C.M.~Group, Cavendish Laboratory, University of Cambridge, Cambridge CB3 0HE, United Kingdom}}
\end{center}

\setcounter{equation}{0}
\setcounter{figure}{0}
\setcounter{table}{0}
\setcounter{page}{1}

\renewcommand{\thesection}{S\arabic{section}}
\renewcommand{\theequation}{S\arabic{equation}}
\renewcommand{\thefigure}{S\arabic{figure}}
\renewcommand{\thetable}{S\arabic{table}}

\section{\label{supply:lswt}Linear Spin Wave Theory}

In the linear spin wave analysis, one first rotates the local coordinate frame defined at each magnetic site such that the $z$ axis align with the spin \cite{Jones_1987}, while the $x$ and $y$ axes can be chosen freely as long as they are orthogonal and $\hat{\mathbf{z}} = \hat{\mathbf{x}} \times \hat{\mathbf{y}}$ \cite{PhysRevResearch.3.013179}. For the polarized state with field angle $\beta$, the axes of the rotated coordinates are defined by $\hat{\mathbf{z}} = \hat{\mathbf{a}} \cos \beta + \hat{\mathbf{b}} \sin \beta $, $\hat{\mathbf{x}} = - \hat{\mathbf{c}}$, $\hat{\mathbf{y}} = - \hat{\mathbf{a}} \sin \beta + \hat{\mathbf{b}} \cos \beta $, where $\hat{\mathbf{a}}$, $\hat{\mathbf{b}}$, and $\hat{\mathbf{c}}$ are unit vectors along the crystallographic axes $a$, $b$, and $c$, respectively, see Fig.~\ref{figure:geometry}. One then performs Holstein-Primakoff transformation \cite{PhysRev.58.1098} to represent the spins in terms of bosons (i.e., magnons), perform a $1/S$ expansion of the Hamiltonian ($\sim S^2$ in the classical limit), and discard terms are of orders lower than $S$. These procedures are well established and described in details elsewhere (see Refs.~\cite{PhysRevB.89.054420,annurev-conmatphys-031620-104715,PhysRevLett.126.147201} for example), so we do not repeat them here. \\

For the $J K\Gamma \Gamma'$ model \eqref{spinmodel} under an in-plane magnetic field $(h_a,h_b,h_c)=h(\cos \beta,\sin \beta,0)$, the linear spin wave Hamiltonian of the polarized state is given by $H = (S/2) \sum_\mathbf{k} \Psi_\mathbf{k}^\dagger \mathcal{H}_\mathbf{k} \Psi_\mathbf{k}$, where $\Psi_\mathbf{k} = (b_{1\mathbf{k}},b_{2 \mathbf{k}},b_{1-\mathbf{k}}^\dagger,b_{2-\mathbf{k}}^\dagger)$ and $\mathcal{H}_\mathbf{k}$ is a four dimensional Hermitian matrix,
\begin{subequations}
\begin{align}
& \mathcal{H}_\mathbf{k} = \begin{pmatrix} \mathcal{A}_\mathbf{k} & \mathcal{B}_\mathbf{k} \\ \mathcal{B}_{-\mathbf{k}}^* & \mathcal{A}_{-\mathbf{k}}^\mathrm{T} \end{pmatrix} , \label{hamiltonianfourier} \\
& \mathcal{A}_\mathbf{k} = \frac{1}{2} \begin{pmatrix} 2 (h - c_1) & c_2 f_\mathbf{k}^* + c_3 + c_4 g_\mathbf{k}^* \\ c_2 f_\mathbf{k} + c_3 + c_4 g_\mathbf{k} & 2 (h - c_1) \end{pmatrix} , \label{Akdefine} \\
& \mathcal{B}_\mathbf{k} = \frac{1}{2} \begin{pmatrix} 0 & c_5 f_\mathbf{k}^* - c_3 - c_4 g_\mathbf{k}^* + 2 i \big[ c_6 (f_\mathbf{k}^* - 3) + c_7 g_\mathbf{k}^* \big] \\ c_5 f_\mathbf{k} - c_3 - c_4 g_\mathbf{k} + 2 i \big[ c_6 (f_\mathbf{k} - 3) + c_7 g_\mathbf{k} \big] & 0 \end{pmatrix} , \label{Bkdefine}
\end{align}
\end{subequations}
where $c_i$ are (real) linear combinations of the couplings, while $f_\mathbf{k}$ and $g_\mathbf{k}$ are functions of the (crystal) momentum, see \eqref{cidefine} and related discussions in the main text. To obtain the linear spin wave dispersion, $\mathcal{H}_\mathbf{k}$ has to be diagonalized by a Bogoliubov transformation $T_\mathbf{k}$ satisfying $T_\mathbf{k} \eta T_\mathbf{k}^\dagger = \eta$, where $\eta = \mathrm{diag} (1,1,-1,-1)$, in order to preserve the commutation relation of bosons. The magnon bands are given by
\begin{equation} \label{eigenvalue}
\omega_\pm (\mathbf{k}) = \frac{S}{2} \sqrt{E(\mathbf{k}) \pm \Delta (\mathbf{k})} ,
\end{equation}
where $E (\mathbf{k})$ and $\Delta (\mathbf{k})$ are defined in \eqref{Ekdefine} and \eqref{Dksquaredefine} in the main text. \\

We consider a stable polarized state, where the excitation gap $\min_\mathbf{k} \omega_- (\mathbf{k})$ is greater than zero. As the field strength $h$ increases, the excitation gap should increase as well, so that the polarized state becomes more stable, rather than undergoing a magnon instability in which the excitation gap vanishes. Based on this physical expectation, we can assume
\begin{equation} \label{stability}
h - c_1 > 0
\end{equation}
always, which is argued as follows. Eq.~\eqref{stability} obviously holds for $c_1 \leq 0$ for all $h$. For $c_1 > 0$, assuming that a finite excitation gap is possible for some $h < c_1$, we can then dial up $h$ such that $h=c_1$. At the $\mathrm{K}$ or $\mathrm{K}'$ point, where $f_\mathbf{k}=0$, we will have $\omega_{\pm} (\mathbf{k}) = \sqrt{- 4 \lvert - 3 c_6 + c_7 g_\mathbf{k} \rvert^2}$, which is either zero or imaginary, neither being physically sensible. Therefore, magnon stability is only consistent with $h - c_1 > 0$. \\

A quick inspection of \eqref{Dksquaredefine} reveals that $\Delta^2 (\mathbf{k})$ consists of a field dependent part and a field independent part. With the assumption \eqref{stability}, we now claim that if $\Delta (\mathbf{k}) = 0$, then it is only physically sensible that $c_2 f_\mathbf{k} + c_3 + c_4 g_\mathbf{k} = 0$, unless the system sits at a critical point where a transition to the polarized state occurs. Suppose that the contrary is true, i.e., $\Delta (\mathbf{k}) = 0$ and $c_2 f_\mathbf{k} + c_3 + c_4 g_\mathbf{k} \neq 0$ at some $h=h_*$, with $h_*-c_1 > 0$. The field dependent part, which is positive, must cancel the field independent part exactly in $\Delta^2 (\mathbf{k})$. Let $h'=h_*-\epsilon$ with $0 < \epsilon < h_* - c_1$. We have $h'-c_1 > 0$, but $\Delta^2 (\mathbf{k}) < 0$ at $h=h'$, i.e., $\omega_\pm (\mathbf{k})$ develops an imaginary component, which is unphysical. Therefore, $h$ cannot be less than $h_*$. In case $h_*$ marks the phase boundary, we can further increase the field to some $h>h_*$ to obtain a stable polarized state. We have thus established \\

\noindent \quad \textbf{Proposition 1.} Under the stability requirement $h-c_1>0$, $c_2 f_\mathbf{k} + c_3 + c_4 g_\mathbf{k} = 0$ is a necessary condition for $\Delta (\mathbf{k}) = 0$. \\

We refer to the gap between the upper and lower magnon bands, $\min_\mathbf{k} [\omega_+ (\mathbf{k}) - \omega_- (\mathbf{k})] \geq 0$, as the \textit{band gap}, which is not to be confused with the excitation gap. From \eqref{eigenvalue}, we see that the band gap vanishes if and only if $\Delta (\mathbf{k}) = 0$ for some $\mathbf{k}=\mathbf{k}_*$. For a fixed set of couplings $\{J, K, \Gamma, \Gamma' \}$, the analysis in the previous two paragraphs implies, within a stable polarized state, \\

\noindent \quad \textbf{Corollary 1.} If the band gap is zero, then it remains zero as $h$ varies;

\noindent \quad \textbf{Corollary 2.} If the band gap is finite, then it remains finite as $h$ varies, unless $h \longrightarrow \infty$ while the couplings stay finite. \\

To see why these observations are useful, we first note that the Chern number is a topological invariant that can never change as long as a finite band gap is maintained. A topological phase transition can only occur when the band gap vanishes. Therefore, within a stable in-plane field polarized state and for a finite $h$, Corollaries 1 and 2 respectively imply \\

\textbf{Lemma 1.} If a topological transition exists, the phase boundary, which must be a parameter region where the band gap goes to zero, is independent of the field strength;

\textbf{Lemma 2.} The Chern number of each magnon band, which is well defined when the band gap is finite, is independent of the field strength. \\

We are now ready to solve analytically for regions in the $J K \Gamma \Gamma'$ parameter space where the band gap vanishes, which are potential phase boundaries for topological transitions. \\

\fbox{$\beta=0$.} With $c_4=0$ and $c_6=0$, \eqref{Dksquaredefine} reads
\begin{equation} \label{Dksquaredefineha}
\Delta^2 (\mathbf{k}) = 16 ( h-c_1 )^2 \big\lvert c_2 f_\mathbf{k} + c_3 \big\rvert^2 + ( c_2+c_5 ) \big[ (c_2 + c_5) c_3^2 ( f_\mathbf{k}^* - f_\mathbf{k} )^2 + 8 c_3 c_7^2 ( g_\mathbf{k} f_\mathbf{k}^* - g_\mathbf{k}^* f_\mathbf{k} ) ( g_\mathbf{k} - g_\mathbf{k}^*) + 4 ( c_2 - c_5 ) c_7^2 ( g_\mathbf{k} f_\mathbf{k}^* - g_\mathbf{k}^* f_\mathbf{k} )^2 \big] .
\end{equation}
The band gap is zero if and only if there is at least one $\mathbf{k}$ such that $\Delta (\mathbf{k}) = 0$. According to Proposition 1, we require
\begin{equation} \label{proposition1ha}
c_2 f_\mathbf{k} + c_3 = 0 ,
\end{equation}
which makes the field dependent part of $\Delta^2 (\mathbf{k})$ zero. The field independent part should be zero as well. We solve these conditions on a case by case basis. \\

\noindent \quad \textbf{Case 1.}~$c_2=0$. We must have $c_3=0$. When $\mathbf{k}=\mathrm{K},\mathrm{K}'$, $f_\mathbf{k} = 0$, and $\Delta^2 (\mathbf{k}) = 0$ is satisfied.

\noindent \quad \textbf{Case 2.}~$c_2 \neq 0$.

\noindent \quad \quad \textbf{Case 2.1.}~$c_3 = 0$. When $\mathbf{k}=\mathrm{K},\mathrm{K}'$, $f_\mathbf{k} = 0$, and $\Delta^2 (\mathbf{k}) = 0$ is satisfied.

\noindent \quad \quad \textbf{Case 2.2.}~$c_3 \neq 0$. The real and imaginary parts of \eqref{proposition1ha} respectively read
\begin{subequations}
\begin{align}
c_2 (1 + \cos k_1 + \cos k_2) + c_3 &= 0 , \label{proposition1hare} \\
c_2 (\sin k_1 + \sin k_2 ) &= 0 . \label{proposition1haim}
\end{align}
\end{subequations}
\eqref{proposition1haim} implies $k_2 = - k_1 + 2 \pi$ or $k_2 = k_1 + \pi$.

\noindent \quad \quad \quad \textbf{Case 2.2.1.}~$k_2 = - k_1 + 2 \pi$. Eq.~\eqref{proposition1hare} becomes $c_2 (1 + 2 \cos k_1) + c_3 = 0$, which only admits a solution when $-1 \leq -c_3/c_2 \leq 3$. In this case, $f_\mathbf{k} = -c_3/c_2$, and \eqref{Dksquaredefineha} becomes
\begin{equation}
\Delta^2 (\mathbf{k}) = \left( \frac{8 c_3}{c_2} \right)^2 c_7^2 (c_2 + c_5)^2 \sin^2 k_1 .
\end{equation}
The first bracket on the right hand side is assumed to be nonzero, so $\Delta^2 (\mathbf{k}) = 0$ if and only if (i) $c_7=0$, (ii) $c_2 + c_5 = 0$, or (iii) $\sin k_1 = 0$, which implies $k_1 = 0, \pi$, which in turn implies $3 c_2 + c_3 = 0$ or $- c_2 + c_3 = 0$, respectively.

\noindent \quad \quad \quad \textbf{Case 2.2.2.}~$k_2 = k_1 + \pi$. Eq.~\eqref{proposition1hare} reads $c_2 + c_3 = 0$. In this case, $f_\mathbf{k}=1$ and $g_\mathbf{k}$ is real at $k_1=0,\pi$, which yield $\Delta^2 (\mathbf{k}) = 0$ by \eqref{Dksquaredefineha}. \\

Collecting all the results, the parameter regions where the band gap vanishes satisfy one of the following equations: (I) $c_3 = 0$, (II) $3 c_2 + c_3 = 0$, (III) $- c_2 + c_3 = 0$, (IV,V) $c_2 + c_3 = 0$, (VI) $c_2 + c_5 = 0$ if $-1 \leq - c_3 / c_2 \leq 3$, and (VII) $c_7=0$ if $-1 \leq - c_3 / c_2 \leq 3$. The reason that we use two labels IV and V for the equation $c_2 + c_3 = 0$ will be clear when we discuss the case $0 < \beta \leq \pi/6$. \\

\fbox{$0 < \beta \leq \pi/6$.} The band gap is zero if and only if there is at least one $\mathbf{k}$ such that $\Delta (\mathbf{k}) = 0$ [see \eqref{Dksquaredefine} in the main text]. According to Proposition 1, we require
\begin{equation} \label{proposition1}
c_2 f_\mathbf{k} + c_3 + c_4 g_\mathbf{k}= 0 ,
\end{equation}
which makes the field dependent part of $\Delta^2 (\mathbf{k})$ zero. The field independent part should be zero as well. We solve these conditions on a case by case basis. \\

\noindent \quad \textbf{Case 1.} $K + 2 \Gamma - 2 \Gamma' = 0$. Then, $c_3 = 0$ and $c_4 = 0$. At $\mathbf{k} = \mathrm{K}, \mathrm{K}'$, $f_\mathbf{k}=0$, and $\Delta^2 (\mathbf{k}) = 0$ from \eqref{Dksquaredefine}.

\noindent \quad \textbf{Case 2.} $K + 2 \Gamma - 2 \Gamma' \neq 0$. Then, $c_3 \neq 0$ and $c_4 \neq 0$. The real and imaginary parts of \eqref{proposition1} respectively read
\begin{subequations}
\begin{align}
( c_2 + c_3 ) + ( c_2 + c_4 ) \cos k_1 + ( c_2 - c_4 ) \cos k_2 &= 0 , \label{proposition1re} \\
( c_2 + c_4 ) \sin k_1 + ( c_2 - c_4 ) \sin k_2 &= 0 . \label{proposition1im}
\end{align}
\end{subequations}
\noindent \quad \quad \textbf{Case 2.1.} $c_2 = c_4$. Eq.~\eqref{proposition1im} implies $k_1 = 0,\pi$. Substituting $k_1 = \pi$ in \eqref{proposition1re} leads to $c_3 - c_4 = 0$, or $K + 2 \Gamma - 2 \Gamma' = 0$, a contradiction. We thus discard $k_1 = \pi$. Substituting $k_1 = 0$ in \eqref{proposition1re} leads to $c_3 + 3 c_4 = 0$, which holds only at $\beta = \pi / 12$. We can then choose $k_2 = 0, \pi$ so that $f_\mathbf{k}$ and $g_\mathbf{k}$ are real, and $\Delta^2 (\mathbf{k}) = 0$ is satisfied.

\noindent \quad \quad \textbf{Case 2.2.} $c_2 = - c_4$. Eq.~\eqref{proposition1im} implies $k_2 = 0 , \pi$. Substituting $k_2 = 0$ in \eqref{proposition1re} leads to $c_3 - 3 c_4 = 0$, or $K + 2 \Gamma - 2 \Gamma' = 0$, a contradiction. We thus discard $k_2 = 0$. Substituting $k_2 = \pi$ in \eqref{proposition1re} leads to $c_3 + c_4 = 0$, which holds only at $\beta = \pi / 6$. We can then choose $k_1 = 0, \pi$ so that $f_\mathbf{k}$ and $g_\mathbf{k}$ are real, and $\Delta^2 (\mathbf{k}) = 0$ is satisfied.

\noindent \quad \quad \textbf{Case 2.3.} $\lvert c_2 \rvert \neq \lvert c_4 \rvert$.

\noindent \quad \quad \quad \textbf{Case 2.3.1.} $(k_1 , k_2) \in \lbrace (0,0), (0,\pi), (\pi,0), (\pi,\pi) \rbrace$. Eq.~\eqref{proposition1im} is satisfied. Eq.~\eqref{proposition1re} implies $3 c_2 + c_3 = 0$, $c_2 + c_3 + 2 c_4 = 0$, $c_2 + c_3 - 2 c_4 = 0$, or $- c_2 + c_3 = 0$, respectively. Since $f_\mathbf{k}$ and $g_\mathbf{k}$ are real, $\Delta^2 (\mathbf{k}) = 0$ is satisfied.

\begin{figure}
\includegraphics[scale=0.35]{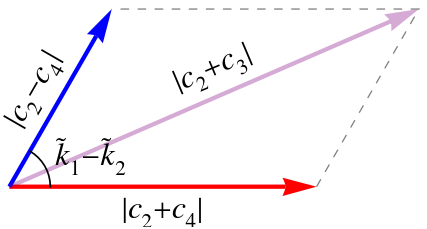}
\caption{\label{figure:vector}Eq.~\eqref{vectorsummation} can be interpreted as the sum of two vectors.}
\end{figure}

\noindent \quad \quad \quad \textbf{Case 2.3.2.} $(k_1 , k_2) \notin \lbrace (0,0), (0,\pi), (\pi,0), (\pi,\pi) \rbrace$. We write \eqref{proposition1re} and \eqref{proposition1im} as
\begin{subequations}
\begin{align}
\lvert c_2 + c_4 \rvert \cos \tilde{k}_1 + \lvert c_2 - c_4 \rvert \cos \tilde{k}_2 &= - ( c_2 + c_3 ) , \label{proposition1reabs} \\
\lvert c_2 + c_4 \rvert \sin \tilde{k}_1 + \lvert c_2 - c_4 \rvert \sin \tilde{k}_2 &= 0 , \label{proposition1imabs}
\end{align}
\end{subequations}
where $\tilde{k}_1 = k_1$ if $c_2 + c_4 > 0$ and $\tilde{k}_1 = k_1 + \pi$ if $c_2 + c_4 < 0$; $\tilde{k}_2$ is defined in a similar way. Squaring both sides of \eqref{proposition1reabs} and \eqref{proposition1imabs}, and adding up the results lead to
\begin{equation} \label{vectorsummation}
\lvert c_2 + c_4 \rvert^2 + \lvert c_2 - c_4 \rvert^2 + 2 \lvert c_2 + c_4 \rvert \lvert c_2 - c_4 \rvert \cos (\tilde{k}_1 - \tilde{k}_2) = \lvert c_2 + c_3 \rvert^2 .
\end{equation}
\eqref{vectorsummation} can be interpreted as a summation of two vectors, one of length $\lvert c_2 + c_4 \rvert$ and the other $\lvert c_2 - c_4 \rvert$, with an angle $\tilde{k}_1 - \tilde{k}_2$ in between, which results in a vector of length $\lvert c_2 + c_3 \rvert$, see Fig.~\ref{figure:vector}. With this interpretation, we deduce that \eqref{vectorsummation} admits a solution when
\begin{equation} \label{solubilitysupplement}
\big\lvert \lvert c_2 + c_4 \rvert - \lvert c_2 - c_4 \rvert \big\rvert \leq \lvert c_2 + c_3 \rvert \leq \lvert c_2 + c_4 \rvert + \lvert c_2 - c_4 \lvert .
\end{equation}
The right (left) equality holds when the two vectors (anti-)align, i.e., $\tilde{k}_1 - \tilde{k}_2 = 0$ ($\pi$), where the resulting vector reaches its longest (shortest) possible length. The equalities in \eqref{solubilitysupplement}, however, require $(k_1 , k_2) \in \lbrace (0,0), (0,\pi), (\pi,0), (\pi,\pi) \rbrace$, which can be seen from \eqref{proposition1reabs}. This is a contradiction, but we note that these momenta have been covered in Case 2.3.1. We can thus focus on the inequalities in \eqref{solubilitysupplement}, and assume that they are satisfied. From \eqref{vectorsummation},
\begin{subequations}
\begin{align}
\cos ( \tilde{k}_1 - \tilde{k}_2 ) &= \frac{ \lvert c_2 + c_3 \rvert^2 - \lvert c_2 + c_4 \rvert^2 - \lvert c_2 - c_4 \rvert^2 }{ 2 \lvert c_2 + c_4 \rvert \lvert c_2 - c_4 \rvert } , \label{cosk1k2} \\
\sin ( \tilde{k}_1 - \tilde{k}_2 ) &= \pm \sqrt{1 - \cos^2 ( \tilde{k}_1 - \tilde{k}_2 )} . \label{sink1k2}
\end{align}
\end{subequations}
Substituting \eqref{proposition1imabs} in \eqref{proposition1reabs} yields
\begin{subequations}
\begin{align}
\sin \tilde{k}_1 & = - \frac{\lvert c_2 - c_4 \rvert}{c_2 + c_3} \sin ( \tilde{k}_1 - \tilde{k}_2 ) , \label{sink1} \\
\sin \tilde{k}_2 &= \frac{\lvert c_2 + c_4 \rvert}{c_2 + c_3} \sin ( \tilde{k}_1 - \tilde{k}_2 ) . \label{sink2}
\end{align}
\end{subequations}
Notice that $c_2+c_3 \neq 0$ due to \eqref{solubilitysupplement}. From \eqref{sink1} and \eqref{sink2},
\begin{subequations}
\begin{align}
& - \sin ( \tilde{k}_1 - \tilde{k}_2 ) \cos \tilde{k}_1 + \cos ( \tilde{k}_1 - \tilde{k}_2 ) \sin \tilde{k}_1 = \frac{ \lvert c_2 + c_4 \rvert }{ c_2 + c_3 } \sin ( \tilde{k}_1 - \tilde{k}_2 ) \iff \cos \tilde{k}_1 = - \frac{ \lvert c_2 + c_4 \rvert }{ c_2 + c_3 } - \frac{ \lvert c_2 - c_4 \rvert }{ c_2 + c_3 } \cos ( \tilde{k}_1 - \tilde{k}_2 ) , \label{cosk1} \\
& \sin ( \tilde{k}_1 - \tilde{k}_2 ) \cos \tilde{k}_2 + \cos ( \tilde{k}_1 - \tilde{k}_2 ) \sin \tilde{k}_2 = - \frac{ \lvert c_2 - c_4 \rvert }{ c_2 + c_3 } \sin ( \tilde{k}_1 - \tilde{k}_2 ) \iff \cos \tilde{k}_2 = - \frac{ \lvert c_2 - c_4 \rvert }{ c_2 + c_3 } - \frac{\lvert c_2 + c_4 \rvert}{c_2 + c_3} \cos ( \tilde{k}_1 - \tilde{k}_2 ) . \label{cosk2}
\end{align}
\end{subequations}
In order for \eqref{sink1}, \eqref{sink2}, \eqref{cosk1}, and \eqref{cosk2} to admit a solution $( \tilde{k}_1, \tilde{k}_2 )$, we need to verify that the absolute values of the right hand sides are less than or equal to unity. We calculate
\begin{subequations}
\begin{align}
& \sin^2 \tilde{k}_1 + \cos^2 \tilde{k}_1 = \frac{ \lvert c_2 - c_4 \rvert^2 + \lvert c_2 + c_4 \rvert^2 + 2 \lvert c_2 + c_4 \rvert \lvert c_2 - c_4 \rvert \cos ( \tilde{k}_1 - \tilde{k}_2 )}{ \lvert c_2 + c_3 \rvert^2 } = 1 , \\
& \sin^2 \tilde{k}_2 + \cos^2 \tilde{k}_2 = \frac{ \lvert c_2 + c_4 \rvert^2 + \lvert c_2 - c_4 \rvert^2 + 2 \lvert c_2 - c_4 \rvert \lvert c_2 + c_4 \rvert \cos ( \tilde{k}_1 - \tilde{k}_2 )}{ \lvert c_2 + c_3 \rvert^2 } = 1 ,
\end{align}
\end{subequations}
where the last equalities follow from \eqref{vectorsummation}. If the squares of two real numbers add up to unity, then each must be less than or equal to unity. We have demonstrated that there indeed exists $( \tilde{k}_1, \tilde{k}_2 )$, defined implicitly via \eqref{sink1}, \eqref{sink2}, \eqref{cosk1}, and \eqref{cosk2} in conjunction with \eqref{cosk1k2} and \eqref{sink1k2}, which solves \eqref{proposition1reabs} and \eqref{proposition1imabs}.

Substituting \eqref{proposition1} in \eqref{Dksquaredefine},
\begin{equation}
\Delta^2 ( \mathbf{k} ) = - \frac{4 ( c_2 + c_5 )^2 ( c_3 c_7 + 3 c_4 c_6 )^2 ( f_\mathbf{k} - f_\mathbf{k}^* )^2 }{c_4^2} = - \frac{ \cos^2 (3 \beta) ( c_2 + c_5 )^2 ( K - \Gamma + \Gamma' )^2 ( K + 2 \Gamma - 2 \Gamma' )^2 ( f_\mathbf{k} - f_\mathbf{k}^* )^2 }{ 6 c_4^2 } .
\end{equation}
If $ \beta = \pi / 6 $, then $ \cos (3 \beta) = 0 $, and $\Delta^2 (\mathbf{k}) = 0$ is satisfied. For $ 0 < \beta < \pi / 6 $, $\cos (3 \beta) \neq 0$. $f_\mathbf{k} - f_\mathbf{k}^*$ vanishes if and only if $f_\mathbf{k}$ is real, but this implies, via \eqref{proposition1im}, $c_4=0$ or $( k_1 , k_2 ) \in \lbrace (0,0), (0,\pi), (\pi,0), (\pi,\pi) \rbrace $, both of which violate the initial assumptions. Thus $f_\mathbf{k} - f_\mathbf{k}^* \neq 0$. Furthermore, $K + 2 \Gamma - 2 \Gamma' \neq 0$ by assumption. Therefore, $\Delta^2 (\mathbf{k}) = 0$ if and only if (i) $c_2 + c_5 = 0$ or (ii) $K - \Gamma + \Gamma' =0$. \\

Collecting all the results, for $ 0 < \beta < \pi/6 $, the parameter regions where the band gap vanishes satisfy one of the following equations: (I) $ K + 2 \Gamma - 2 \Gamma' = 0 $, (II) $ 3 c_2 + c_3 = 0$, (III) $ - c_2 + c_3 = 0 $, (IV) $ c_2 + c_3 + 2 c_4 = 0 $, (V) $ c_2 + c_3 - 2 c_4 = 0 $, (VI) $ c_2 + c_5 = 0 $ if \eqref{solubilitysupplement} holds, and (VII) $ K - \Gamma + \Gamma' = 0 $ if \eqref{solubilitysupplement} holds. Setting $ \beta=0 $, these criteria are equivalent to those solved earlier for $\beta = 0$, which was the reason that we used two labels IV and V for $ c_2 + c_3 = 0 $, as $c_4 = 0$ there. On the other hand, for $ \beta = \pi / 6 $, the gap is zero whenever (I) or \eqref{solubilitysupplement} holds. (I-VII) are expressed in terms of the couplings and the field angle in Table \ref{table:gapless}.

\section{\label{supply:chern}Computation of Chern number}

The Berry curvature of the $n^\mathrm{th}$ magnon band at the momentum $\mathbf{k}$ is defined in terms of the Bogoliubov transformation $T_\mathbf{k}$ as \cite{PhysRevB.89.054420}
\begin{equation} \label{berrycurvature}
\Omega_{n\mathbf{k}} = i \epsilon_{\mu \nu} \left ( \eta \frac{ \partial T_\mathbf{k}^\dagger }{ \partial k_\mu} \eta \frac{ \partial T_\mathbf{k} }{ \partial k_\nu } \right )_{nn} ,
\end{equation}
where $\epsilon$ is the totally antisymmetric tensor and $\mu, \nu \in \lbrace x, y \rbrace$. (Caution: The expression within the brackets on the right hand side is a matrix, and the subscript $nn$ means the entry at the $n^\mathrm{th}$ row and the $n^\mathrm{th}$ column; $n$ is not a dummy index that is being summed over.) Integrating \eqref{berrycurvature} over the magnetic Brillouin zone gives the Chern number of the $n^\mathrm{th}$ band, 
\begin{equation} \label{chernnumber}
\nu_n = \frac{1}{2 \pi} \int_\mathrm{MBZ} \mathrm{d} k_x \mathrm{d} k_y \, \Omega_{n \mathbf{k}} .
\end{equation}

This section explains the method that we use to compute the Chern number in a discretized Brillouin zone, which was introduced in Ref.~\cite{JPSJ.74.1674} and based on a $U(1)$ lattice gauge theory (see also Refs.~\cite{PhysRevB.98.060405,s12200-019-0963-9}). The Berry curvature \eqref{berrycurvature} multiplied by the integral measure, $\Omega_{n \mathbf{k}} \, \mathrm{d} k_x \mathrm{d} k_y $, is invariant under a coordinate transformation \cite{vanderbilttextbook}, e.g.
\begin{subequations}
\begin{align}
& \Omega_n ( k_x , k_y ) \, \mathrm{d} k_x \mathrm{d} k_y = F_n ( k_1 , k_2 ) \, \mathrm{d} k_1 \mathrm{d} k_2 , \label{berrycurvatureinvariance} \\
& F_n ( k_1 , k_2 ) \equiv i \left ( \eta \frac{ \partial T_\mathbf{k}^\dagger }{ \partial k_1} \eta \frac{ \partial T_\mathbf{k} }{ \partial k_2 } - \eta \frac{ \partial T_\mathbf{k}^\dagger }{ \partial k_2} \eta \frac{ \partial T_\mathbf{k} }{ \partial k_1 } \right)_{nn} . \label{berrycurvature12}
\end{align}
\end{subequations}
(More formally, the differential 2-form $\Omega_{n \mathbf{k}} \, \mathrm{d} k_x \wedge \mathrm{d} k_y$ is coordinate independent.) Let $\mathcal{N}$ be the number of sites per magnetic unit cell, in particular $\mathcal{N}=2$ for the polarized state. If there is a finite magnon pairing term, then the $2 \mathcal{N}$ dimensional Hamiltonian matrix $\mathcal{H}_\mathbf{k}$ has a particle-hole redundancy by construction. The columns of the $2 \mathcal{N}$ dimensional Bogoliubov transformation matrix $T_\mathbf{k}$ are arranged such that the first (last) $\mathcal{N}$ columns belong to the particle (hole) sector. \\

Let $\lvert n (\mathbf{k}) \rangle \equiv ( \mathbf{u}_n (\mathbf{k}), \mathbf{v}_n (\mathbf{k}) )$ be the $n^\mathrm{th}$ vector of $T_\mathbf{k}$. We have introduced the $\mathcal{N}$ dimensional vector $\mathbf{u}_n ( \mathbf{k} )$ [$\mathbf{v}_n ( \mathbf{k} )$] as the first [second] half of $ \lvert n (\mathbf{k}) \rangle $. In the rest of this section, we focus on the particle sector, i.e., $1 \leq n \leq \mathcal{N}$. Next, we define the Berry connection of the $n^\mathrm{th}$ band at the momentum $\mathbf{k}$ as
\begin{equation} \label{berryconnection}
A_{n , \lambda} ( \mathbf{k} ) = i \langle n ( \mathbf{k} ) \rvert \eta \partial_{k_\lambda} \lvert n ( \mathbf{k} ) \rangle = i \left[ \mathbf{u}_n^\dagger ( \mathbf{k} ) \partial_{k_\lambda} \mathbf{u}_n ( \mathbf{k} ) - \mathbf{v}_n^\dagger ( \mathbf{k} ) \partial_{k_\lambda} \mathbf{v}_n ( \mathbf{k} ) \right] ,
\end{equation}
where $\lambda=1,2$. Since $\langle n ( \mathbf{k} ) \rvert \eta \lvert n ( \mathbf{k} ) \rangle = 1$, $\langle n ( \mathbf{k} ) \rvert \eta \partial_{k_\lambda} \lvert n ( \mathbf{k} ) \rangle$ is purely imaginary and hence $A_{n , \lambda}$ is purely real. Using the definition \eqref{berrycurvature12}, it can be straightforwardly verified that
\begin{equation} \label{berrycurvaturefromconnection}
F_n ( \mathbf{k} ) = \left[ \partial_{k_1} A_{n , 2} (\mathbf{k}) - \partial_{k_2} A_{n , 1} (\mathbf{k}) \right] .
\end{equation}

On the discretized Brillouin zone, suppose that the spacings of momenta along the $k_1$ and $k_2$ directions are $\delta k_1$ and $\delta k_2$, respectively. If we make $\delta k_\lambda$ small enough, we can approximate \eqref{berryconnection} and \eqref{berrycurvaturefromconnection} as
\begin{subequations}
\begin{align}
A_{n , \lambda} \, \delta k_\lambda &\approx i \left[ \langle n ( \mathbf{k} ) \lvert \eta \rvert n ( \mathbf{k} + \delta k_\lambda \hat{\mathbf{\lambda}} ) \rangle - 1 \right] , \\
F_n ( \mathbf{k} ) \, \delta k_1 \delta k_2 &\approx \left [A_{n , 2} ( \mathbf{k} + \delta k_1 \hat{1} ) - A_{n , 2} ( \mathbf{k} ) \right] \delta k_2 - \left [A_{n , 1} ( \mathbf{k} + \delta k_2 \hat{2} ) - A_{n , 1} ( \mathbf{k} ) \right] \delta k_1 . \label{berrycurvatureapproximate}
\end{align}
\end{subequations}
We define the $U(1)$ link variable
\begin{equation} \label{linkvariable}
U_\lambda ( \mathbf{k} ) = \frac{ \langle n ( \mathbf{k} ) \rvert \eta \lvert n ( \mathbf{k} + \delta k_\lambda \hat{\lambda} ) \rangle }{ \left\lvert \langle n ( \mathbf{k} ) \rvert \eta \lvert n ( \mathbf{k} + \delta k_\lambda \hat{\lambda} ) \rangle \right\rvert } = \frac{ \langle n ( \mathbf{k} ) \rvert \eta \lvert n ( \mathbf{k} + \delta k_\lambda \hat{\lambda} ) \rangle }{ 1 + O (\delta k_\lambda^2) } \approx \exp \left[ - i A_{n , \lambda} (\mathbf{k}) \delta k_\lambda \right] , 
\end{equation}
where, to obtain the second equality, we have expanded $\langle n (\mathbf{k}) \rvert \eta \lvert n (\mathbf{k} + \delta k_\lambda \hat{\lambda}) \rangle \approx 1 + \langle n ( \mathbf{k} ) \rvert \eta \partial_{k_\lambda} \lvert n ( \mathbf{k} ) \rangle \delta k_\lambda$ and used the fact that $\langle n ( \mathbf{k} ) \rvert \eta \partial_{k_\lambda} \lvert n ( \mathbf{k} ) \rangle$ is imaginary. Eq.~\eqref{berrycurvatureapproximate} can be expressed in terms of \eqref{linkvariable} as
\begin{equation} \label{berrycurvaturediscrete}
F_n ( \mathbf{k} ) \delta k_1 \delta k_2 \approx i \ln \left[ U_1 ( \mathbf{k} ) U_2 ( \mathbf{k} + \delta k_1 \hat{1} ) U_1^{-1} ( \mathbf{k} + \delta k_2 \hat{2} ) U_2^{-1} ( \mathbf{k} ) \right] .
\end{equation}
Finally, the Chern number of the $n^\mathrm{th}$ magnon band \eqref{chernnumber} is calculated as
\begin{equation}
\nu_n \approx \frac{1}{2 \pi} \sum_{\mathbf{k} \in \mathrm{MBZ}} F_n ( \mathbf{k} ) \, \delta k_1 \delta k_2 .
\end{equation}

The main advantage of using \eqref{berrycurvaturediscrete} over \eqref{berrycurvature12} for computing Chern numbers is that the former is manifestly gauge invariant, i.e., it is unaffected by $\lvert n ( \mathbf{k} ) \rangle \longrightarrow \exp [ - i \chi ( \mathbf{k} ) ] \lvert n ( \mathbf{k} ) \rangle $ as desired, while the latter requires explicit gauge fixings when taking the differences of $T_\mathbf{k}$ to approximate the derivatives. \\

We mention in passing that the thermal Hall conductivity \eqref{thermalhallconductivity} can also be calculated within this framework,
\begin{equation}
\kappa_{xy}^\mathrm{2D} \approx - \frac{k_\mathrm{B}^2 T}{(2 \pi)^2 \hbar} \sum_{n=1}^\mathcal{N} \sum_{\mathbf{k} \in \mathrm{MBZ}} c_2 \left[ g \left( \frac{\hbar \omega_{n \mathbf{k}}}{k_\mathrm{B} T} \right) \right] F_n ( \mathbf{k} ) \, \delta k_1 \delta k_2 ,
\end{equation}
with $F_n ( \mathbf{k} )$ given in \eqref{berrycurvaturediscrete}.

\section{Symmetries}

In this section, we discuss how topological phase diagrams for different in-plane field directions are related by symmetries of the $J K \Gamma \Gamma'$ model, which, in the absence of an external magnetic field, includes a time reversal $\mathcal{T}$ symmetry, a $C_3$ symmetry about the $c$ axis, and a $C_2$ symmetry about the $b$ axis. Let $\mathbf{J} = ( J, K, \Gamma, \Gamma' )$. The Hamiltonian matrix \eqref{hamiltonianfourier} is a function of the parameter $\mathbf{J}$, the field $\mathbf{h}$, and the momentum $\mathbf{k}$, so we write it as $ \mathcal{H}_\mathbf{k} ( \mathbf{J}, h , \beta ) $. \\

Consider a $C_2$ rotation of the field, i.e., $ \beta = \pi/2 - \tilde{\beta} \longrightarrow \pi/2 + \tilde{\beta} $, at a fixed parameter $\mathbf{J}$. Under $C_2$, we also have the mapping $k_1 \longleftrightarrow k_2$, or equivalently $ k_x \longrightarrow -k_x $ and $ k_y \longrightarrow k_y $. By this observation or by explicit calculation \cite{PhysRevLett.126.147201}, one can show that $\mathcal{H}_{ ( k_x , k_y ) } ( \mathbf{J} , h , \pi/2 + \tilde{\beta} ) = \mathcal{H}_{ ( - k_x , k_y ) } ( \mathbf{J} , h , \pi/2 - \tilde{\beta} ) $. As a consequence, if the band gap is zero (finite) at $( \mathbf{J} , h , \pi/2 + \tilde{\beta} )$, then it is also zero (finite) at $( \mathbf{J} , h , \pi/2 - \tilde{\beta} )$. Assume that the band gap is finite. The Bogoliubov transformations are related by $T_{ ( k_x , k_y ) } ( \mathbf{J} , h , \pi/2 + \tilde{\beta} ) = T_{ ( - k_x , k_y ) } ( \mathbf{J} , h , \pi/2 - \tilde{\beta} ) $. By \eqref{berrycurvature} and \eqref{chernnumber}, we have $\Omega_{ n ( k_x , k_y ) } ( \mathbf{J} , h , \pi/2 + \tilde{\beta} ) = - \Omega_{ n ( - k_x , k_y ) } ( \mathbf{J} , h , \pi/2 - \tilde{\beta} ) $ and $\nu_n ( \mathbf{J}, \pi/2 + \tilde{\beta} ) = - \nu_n ( \mathbf{J} , \pi/2 - \tilde{\beta} )$ for the Berry curvatures and the Chern numbers. In other words, the Chern number of the $n^\mathrm{th}$ magnon band in the polarized state switches sign when the field direction is changed to its $C_2$ counterpart. Theorems 1 and 2 in Ref.~\cite{PhysRevLett.126.147201} follow from this. \\

Consider a $C_3$ rotation of the field, i.e., $\beta \longrightarrow \beta + 2 \pi / 3$, at a fixed parameter $\mathbf{J}$. This amounts to holding the field fixed in space while cyclically permuting the $x$, $y$, and $z$ spin components. Therefore, $C_3$ preserves the band gap (which can be either zero or finite) and the Chern number. \\

Consider the action of time reversal $\mathcal{T}$, i.e., $ \mathbf{h} \longrightarrow - \mathbf{h} $ or, specifically for in-plane fields, $ \beta \longrightarrow \beta + \pi $, at a fixed parameter $\mathbf{J}$. We have $ \mathcal{H}_\mathbf{k} ( \mathbf{J} , h , \beta + \pi ) = \mathcal{H}_{-\mathbf{k}}^* ( \mathbf{J} , h , \beta ) $ by a theorem proved in Ref.~\cite{PhysRevB.103.174402}. Therefore, $\mathcal{T}$ preserves the band gap (which can be either zero or finite) but flips the sign of the Chern number. \\

Finally, consider flipping the signs of all couplings, i.e., $\mathbf{J} \longrightarrow -\mathbf{J}$, at a fixed field direction $\beta$. Following Ref.~\cite{PhysRevB.98.060404}, we introduce
\begin{equation}
U = \begin{pmatrix} 0 & 1 & 0 & 0 \\ -1 & 0 & 0 & 0 \\ 0 & 0 & 0 & 1 \\ 0 & 0 & -1 & 0 \end{pmatrix} = \mathbf{1} \otimes ( i \sigma_2 ) ,
\end{equation}
where $\mathbf{1}$ is the two dimensional identity matrix and $\sigma_{1,2,3}$ are the Pauli matrices. We then carry out the transformation $\mathcal{H}_\mathbf{k} ( \mathbf{J} , h , \beta ) \longrightarrow U \mathcal{H}_\mathbf{k} ( \mathbf{J} , h , \beta ) U^\dagger \equiv \tilde{\mathcal{H}}_\mathbf{k} (\mathbf{J} , h , \beta )$, $\Psi_\mathbf{k} \longrightarrow U \Psi_\mathbf{k} \equiv \tilde{\Psi}_\mathbf{k}$, which leaves the Hamiltonian $H = (S/2) \sum_\mathbf{k} \Psi_\mathbf{k}^\dagger \mathcal{H}_\mathbf{k} ( \mathbf{J} , h , \beta ) \Psi_\mathbf{k}$ invariant. The matrix $U$ is unitary, so it preserves the hermicity of $\mathcal{H}_\mathbf{k} ( \mathbf{J}, h , \beta )$. In addition, $U$ preserves the bosonic commutation relation, i.e., $\tilde{\Psi}_\mathbf{k}$ obeys the same commutation rule as $\Psi_\mathbf{k}$, which can be seen from
\begin{equation}
U \eta U^\dagger = ( \mathbf{1} \otimes \sigma_2 )  ( \sigma_3 \otimes \mathbf{1} )  ( \mathbf{1} \otimes \sigma_2 ) = \eta .
\end{equation}
Importantly, it can be shown that $ \mathcal{H}_\mathbf{k} ( -\mathbf{J} , h , \beta ) = \tilde{\mathcal{H}}_{-\mathbf{k}} ( \mathbf{J} , h + 2 c_1 , \beta ) $, where $ c_1 = 3 J + K - \Gamma - 2 \Gamma' $ as defined in \eqref{cidefine}. We can always choose a sufficiently large $h$ such that both $( -\mathbf{J} , h , \beta )$ and $( \mathbf{J}, h + 2 c_1 , \beta )$ yield a stable polarized state. Let $ T_\mathbf{k} ( \mathbf{J} , h , \beta ) $ and $ \tilde{T}_\mathbf{k} (\mathbf{J} , h , \beta ) $ be the Bogoliubov transformations that diagonalize $ \mathcal{H}_\mathbf{k} ( \mathbf{J} , h , \beta ) $ and $ \tilde{\mathcal{H}}_\mathbf{k} ( \mathbf{J} , h , \beta ) $, respectively, which are related by $ T_\mathbf{k} ( - \mathbf{J} , h , \beta ) = \tilde{T}_{-\mathbf{k}} (\mathbf{J} , h + 2 c_1 , \beta ) $. If the band gap is finite at the parameter $\mathbf{J}$, by \eqref{berrycurvature} and \eqref{chernnumber}, we have $\Omega_{n \mathbf{k}} ( -\mathbf{J} , h , \beta ) = \tilde{\Omega}_{n -\mathbf{k}} ( \mathbf{J} , h + 2 c_1 , \beta )$ and $\nu_n ( -\mathbf{J} , h , \beta ) = \nu_n  ( \mathbf{J} , h + 2 c_1 , \beta ) $ for the Berry curvatures and the Chern numbers. By Corollary 2, the Chern number of the $n^\mathrm{th}$ magnon band in the polarized state at $-\mathbf{J}$ is same as that at $\mathbf{J}$. On the other hand, if the band gap vanishes at $\mathbf{J}$, then it also vanishes at $-\mathbf{J}$ by Corollary 1.

\section{Schrieffer-Wolff Transformation}

When the field strength far exceeds the interaction energy scale, we have $E (\mathbf{k}) \sim h^2$ and $\Delta (\mathbf{k}) \sim h$, see \eqref{Ekdefine} and \eqref{Dksquaredefine}. The Schrieffer-Wolff transformation discussed in the main text is given by \cite{PhysRevB.98.060404}
\begin{subequations}
\begin{align}
& H \longrightarrow e^{W} H e^{-W} = H + [W , H] + \ldots , \\
& W = \sum_\mathbf{k} \Psi_\mathbf{k}^\dagger \mathcal{W}_\mathbf{k} \Psi_\mathbf{k}, \quad \mathcal{W}_\mathbf{k} = \frac{1}{2h} \begin{pmatrix} 0 & \mathcal{B}_\mathbf{k} \\ - \mathcal{B}_\mathbf{k}^\dagger & 0 \end{pmatrix} ,
\end{align}
\end{subequations}
which eliminates magnon pairings up to $O (1/h)$, and absorbs their effects in a pure hopping model. From
\begin{equation}
[ W , H ] = \frac{S}{2} \sum_\mathbf{k} \Psi_\mathbf{k}^\dagger \left( \mathcal{W}_\mathbf{k} \, \eta \, \mathcal{H}_\mathbf{k} - \mathcal{H}_\mathbf{k} \, \eta \, \mathcal{W}_\mathbf{k} \right) \Psi_\mathbf{k} ,
\end{equation}
we obtain
\begin{subequations}
\begin{align}
& \mathcal{A}_\mathbf{k} \longrightarrow \mathcal{A}_\mathbf{k} - \frac{1}{h} \mathcal{B}_\mathbf{k} \mathcal{B}_\mathbf{k}^\dagger = \begin{pmatrix} h - c_1 - \lvert b_{-\mathbf{k}} \rvert^2 / 4 h & a_{-\mathbf{k}} / 2 \\ a_\mathbf{k} / 2 & h - c_1 - \lvert b_\mathbf{k} \rvert^2 / 4 h \end{pmatrix} \equiv \mathcal{A}_\mathbf{k}^\mathrm{eff} , \\
& \mathcal{B}_\mathbf{k} \longrightarrow \mathcal{B}_\mathbf{k} - \frac{1}{2 h} \left( \mathcal{B}_\mathbf{k} \mathcal{A}_\mathbf{k} + \mathcal{A}_\mathbf{k} \mathcal{B}_\mathbf{k} \right) = - \frac{1}{8h} \begin{pmatrix} a_\mathbf{k} b_{-\mathbf{k}} + a_{-\mathbf{k}} b_\mathbf{k} & - 4 c_1 b_{-\mathbf{k}} \\ - 4 c_1 b_\mathbf{k} & a_\mathbf{k} b_{-\mathbf{k}} + a_{-\mathbf{k}} b_\mathbf{k} \end{pmatrix} \equiv \mathcal{B}^\mathrm{eff}_\mathbf{k} , \\
& a_\mathbf{k}  \equiv c_2 f_\mathbf{k} + c_3 + c_4 g_\mathbf{k}, b_\mathbf{k} \equiv c_5 f_\mathbf{k} - c_3 - c_4 g_\mathbf{k} + 2 i \left[ c_6 ( f_\mathbf{k} - 3 ) + c_7 g_\mathbf{k} \right] ,
\end{align}
\end{subequations}
and the effective Hamiltonian $\mathcal{H}_\mathbf{k}^\mathrm{eff}$ is given by \eqref{hamiltonianfourier} with $\mathcal{A}_\mathbf{k}$ and $\mathcal{B}_\mathbf{k}$ replaced by $\mathcal{A}_\mathbf{k}^\mathrm{eff}$ and $\mathcal{B}_\mathbf{k}^\mathrm{eff}$, respectively. We can further neglect $\mathcal{B}_\mathbf{k}$, as argued in the following. The linear spin wave dispersion $ \omega_\pm^\mathrm{eff} ( \mathbf{k} ) $ satisfies the characteristic polynomial of $\eta \, \mathcal{H}_\mathbf{k}^\mathrm{eff}$, which has the form
\begin{equation} \label{characteristicpolynomial}
\lambda^4 + ( c_\mathcal{A} + c_\mathcal{B} ) \lambda^2 + c_\mathrm{mix} = 0 .
\end{equation}
The explicit expressions of $c_\mathcal{A}$, $c_\mathcal{B}$, and $c_\mathrm{mix}$ are omitted here for simplicity; we merely note that $c_\mathcal{A} \sim h^2$ depends only on the matrix elements of $\mathcal{A}_\mathbf{k}^\mathrm{eff}$, $ c_\mathcal{B} \sim 1 / h^2 $ only on those of $\mathcal{B}_\mathbf{k}^\mathrm{eff}$, and $c_\mathrm{mix}$ on both. Eq.~\eqref{characteristicpolynomial} is solved by
\begin{equation} \label{quadraticformula}
\lambda^2 = \frac{- ( c_\mathcal{A} + c_\mathcal{B} ) \pm \sqrt{ ( c_\mathcal{A} + c_\mathcal{B} )^2 - 4 c_\mathrm{mix} } }{2} .
\end{equation}
One finds by explicit calculation that the contribution of $\mathcal{B}_\mathbf{k}^\mathrm{eff}$ to the square root in \eqref{quadraticformula} is at best $O ( 1 / \sqrt{h} )$: it can be of higher order, but not lower. From $\omega_\pm^\mathrm{eff} \sim \sqrt{\lambda^2}$, one can perform a large $h$ expansion and deduce that $\mathcal{B}_\mathbf{k}^\mathrm{eff}$ only contributes at $O ( 1 / h^{3/2} )$ and beyond to $\omega_\mathbf{k}^\mathrm{eff}$. Discarding $\mathcal{B}_\mathbf{k}^\mathrm{eff}$ is thus justified, and we are left with a pure hopping Hamiltonian $\mathcal{A}_\mathbf{k}^\mathrm{eff}$. The magnon excitation energies calculated from $\mathcal{A}_\mathbf{k}^\mathrm{eff}$ are equal to those calculated from $\mathcal{H}_\mathbf{k}^\mathrm{eff}$ up to $ O ( 1 / h ) $. \\

Since the effective Hamiltonian $\mathcal{A}_\mathbf{k}^\mathrm{eff}$ is a $2 \times 2$ hermitian matrix, it can be expressed as
\begin{subequations}
\begin{align}
\mathcal{A}_\mathbf{k}^\mathrm{eff} &= d_0 (\mathbf{k}) \mathbf{1} + \mathbf{d} ( \mathbf{k} ) \cdot \bm{\sigma}, \\
d_1 ( \mathbf{k} ) &= \frac{1}{2} \big[ ( c_2 + c_3 ) + ( c_2 + c_4 ) \cos k_1 + ( c_2 - c_4 ) \cos k_2 \big] , \label{d1define} \\
d_2 ( \mathbf{k} ) &= \frac{1}{2} \big[ ( c_2 + c_4 ) \sin k_1 + ( c_2 - c_4 ) \sin k_2 \big] , \label{d2define} \\
\begin{split}
d_3 ( \mathbf{k} ) &= \frac{1}{h} \bigg\lbrace \big[ ( c_5 - c_3 ) + ( c_5 - c_4 ) \cos k_1 + ( c_5 + c_4 ) \cos k_2 \big] \big[ ( c_6 + c_7 ) \sin k_1 + ( c_6 - c_7 ) \sin k_2 \big]  \\
& \quad + \big[ ( c_5 - c_4 ) \sin k_1 + ( c_5 + c_4 ) \sin k_2 \big] \big[ 2 c_6 - ( c_6 + c_7 ) \cos k_1 - ( c_6 - c_7 ) \cos k_2 \big] \bigg\rbrace , \label{d3define}
\end{split}
\end{align}
\end{subequations}
where the explicit expression of $d_0 ( \mathbf{k} )$ is omitted as it does not play a role in the band topology. When $ \mathbf{d} ( \mathbf{k} ) \neq \mathbf{0} $, the Chern number of the lower magnon band is given by \eqref{winding} in the main text. If one of the components of $\mathbf{d} ( \mathbf{k} )$ vanishes throughout the Brillouin zone, then the triple product on the right hand side of \eqref{winding} is identically zero, and consequently the Chern number is zero as well. This provides a sufficient condition for topologically trivial magnons. \\

The parameter regions with $\nu = 0$ in our phase diagrams Figs.~\ref{figure:phasedeg000N}-\ref{figure:phasedeg225S} can now be understood as being related to the vanishing of $d_3 (\mathbf{k})$. We first note from \eqref{cidefine} that both $c_6$ and $c_7$ contain the factor $K - \Gamma + \Gamma'$. Therefore, if $K - \Gamma + \Gamma'=0$, then $c_6=0$ and $c_7=0$, which in turn imply $d_3 ( \mathbf{k} ) = 0$ for all $\mathbf{k}$ by \eqref{d3define}. The rest of the argument is contained in the main text. \\

\begin{figure}
\includegraphics[scale=0.34]{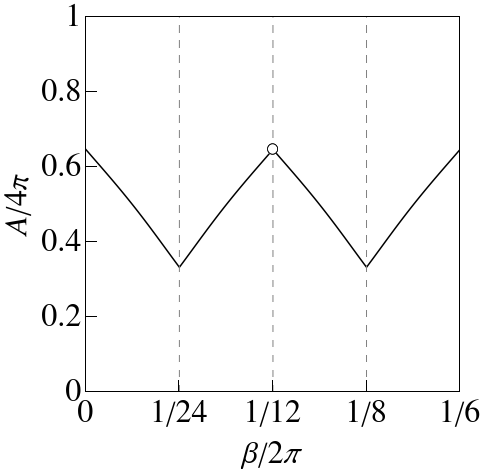}
\caption{\label{figure:area}The total area $A$ of the parameter regions with topological magnons on the $K \Gamma \Gamma'$ sphere, as a function of the field angle $\beta$, over the range $0 \leq \beta \leq \pi / 3$. There is a discontinuity at $\beta = \pi / 6$ (indicated by the empty circle) where $A=0$ is enforced by a $C_2$ symmetry, while large swathes of the parameter space become critical, see Fig.~\ref{figure:phasedeg300S} in the main text.}
\end{figure}

Let $A$ be the total area of the parameter regions on the spherical surface $K^2 + \Gamma^2 + \Gamma'^2 = 1$ with topological magnons. Fig.~\ref{figure:area} shows the dependence of $A$ on the field angle $\beta$. One finds that $A=A_\mathrm{max}$ is maximal at $\beta = 0$. As $\beta$ increases, $A$ first decreases and reaches a local minimum at $\beta = \pi / 12$, then increases again and approaches $A_\mathrm{max}$ as $\beta \longrightarrow \pi / 6$. There is a discontinuity at $\beta = \pi / 6$ as $A$ is forced to $0$ by symmetry.

\section{Third Nearest Neighbor Heisenberg Interaction}

Some of the proposed parametrizations for Kitaev magnets in the literature contain a non-negligible third nearest neighbor Heisenberg interaction, $J_3 \sum_{\langle \langle \langle ij \rangle \rangle \rangle} \mathbf{S}_i \cdot \mathbf{S}_j$, see Refs.~\cite{PhysRevB.93.214431,s41467-017-01177-0} for example. In this section, we add a finite $J_3$ term to the $J K \Gamma \Gamma'$ model \eqref{spinmodel} under an in-plane magnetic field, perform a linear spin wave analysis of the polarized state, and explore its implications on the magnon band topology. \\

We first define $c_1=3 J + K - \Gamma - 2 \Gamma' + 3 J_3$, $c_8 = 2 J_3$, and $t_\mathbf{k}=\exp[i (k_2 - k_1)] + \exp[i (k_1 - k_2)] + \exp[i (k_1 + k_2)]$, while retaining the same definitions of $c_2, \ldots, c_7$ and $f_\mathbf{k}, g_\mathbf{k}$ as before, see \eqref{cidefine} and the succeeding sentence. Then, the Hamiltonian matrix $\mathcal{H}_\mathbf{k}$ assumes the form \eqref{hamiltonianfourier} with
\begin{equation} \label{Akredefine}
\mathcal{A}_\mathbf{k} = \frac{1}{2} \begin{pmatrix} 2 (h - c_1) & c_2 f_\mathbf{k}^* + c_8 t_\mathbf{k}^* + c_3 + c_4 g_\mathbf{k}^* \\ c_2 f_\mathbf{k} + c_8 t_\mathbf{k} + c_3 + c_4 g_\mathbf{k} & 2 (h - c_1) \end{pmatrix}
\end{equation}
and $\mathcal{B}_\mathbf{k}$ same as in \eqref{Bkdefine}. The magnon dispersion is given by  $\omega_\pm (\mathbf{k}) = S \sqrt{E (\mathbf{k}) + \Delta (\mathbf{k})} / 2$, where
\begin{subequations}
\begin{align}
\begin{split} \label{Ekredefine}
& E (\mathbf{k}) = 4 (h-c_1)^2 + \big\lvert c_2 f_\mathbf{k} + c_8 t_\mathbf{k} + c_4 g_\mathbf{k} \big\rvert^2 - \big\lvert c_5 f_\mathbf{k} - c_4 g_\mathbf{k} \big\rvert^2 - 4 \big\lvert c_6 (f_\mathbf{k} - 3) + c_7 g_\mathbf{k} \big\rvert^2 \\ 
&  \qquad \quad + 2 c_3 \big\{ (c_2 + c_5) \mathrm{Re} \big[ f_\mathbf{k} \big] + c_8 \mathrm{Re} \big[ t_\mathbf{k} \big] \big\} ,
\end{split} \\
\begin{split} \label{Dksquareredefine}
& \Delta^2 (\mathbf{k}) = 16 (h - c_1)^2 \big\lvert c_2 f_\mathbf{k} + c_8 t_\mathbf{k} + c_3 + c_4 g_\mathbf{k} \big\rvert^2 - 4 \big\{ \mathrm{Im} \big[ (c_2 f_\mathbf{k} + c_8 t_\mathbf{k} + c_5 f_\mathbf{k}) (c_3 + c_4 g_\mathbf{k}^*) + c_5 c_8 f_\mathbf{k} t_\mathbf{k}^* \big] \big\}^2 \\
& \qquad \quad \; \: - 32 \, \mathrm{Im} \big[ (c_2 f_\mathbf{k} + c_8 t_\mathbf{k} + c_5 f_\mathbf{k}) (3 c_6 - c_7 g_\mathbf{k}^*) + c_6 c_8 f_\mathbf{k} t_\mathbf{k}^* \big] \mathrm{Im} \big[ c_6 f_\mathbf{k} (c_3 + c_4 g_\mathbf{k}^* ) + (c_3 c_7 + 3 c_4 c_6) g_\mathbf{k} \big] \\
& \qquad \quad \; \: + 16 \big\{ \mathrm{Im} \big[ c_5 f_\mathbf{k} (3 c_6 - c_7 g_\mathbf{k}^*) \big] \big\}^2 - 16 \big\{ \mathrm{Im} \big[ (c_2 f_\mathbf{k} + c_8 t_\mathbf{k}) (3 c_6 - c_7 g_\mathbf{k}^*)  + c_6 c_8 f_\mathbf{k} t_\mathbf{k}^* \big] \big\}^2 .
\end{split}
\end{align}
\end{subequations}
$\mathrm{Re} [\ldots]$ and $\mathrm{Im} [\ldots]$ denote the real and imaginary parts of their respective arguments. \\

Making use of the fact that both $f_\mathbf{k}$ and $t_\mathbf{k}$ are zero at $\mathbf{k}=\mathrm{K},\mathrm{K}'$, we can follow the same reasoning as in Sec.~\ref{supply:lswt} to show that magnon stability is only consistent with $h-c_1>0$. We then arrive at Proposition 1, which is modified in the presence of the third nearest neighbor Heisenberg interaction as follows: \\

Under the stability requirement $h-c_1 > 0$, $c_2 f_\mathbf{k} + c_8 t_\mathbf{k} + c_3 + c_4 g_\mathbf{k} = 0$ is a necessary condition for $\Delta (\mathbf{k}) = 0$. \\

Within a stable polarized state, for a fixed set of couplings $\lbrace J, K, \Gamma, \Gamma', J_3 \rbrace$, the modified Proposition 1 implies Corollaries 1 and 2, which in turn imply Lemmas 1 and 2, as in Sec.~\ref{supply:lswt}. In other words, upon including $J_3$, we still have the general property that the topological phase diagrams are independent of the field strength. However, solving for the critical regions where the band gap vanishes is no longer analytically tractable for generic model parameters and field angles. \\

\begin{figure}
\subfloat[]{\label{figure:parameter3}
\includegraphics[scale=0.3]{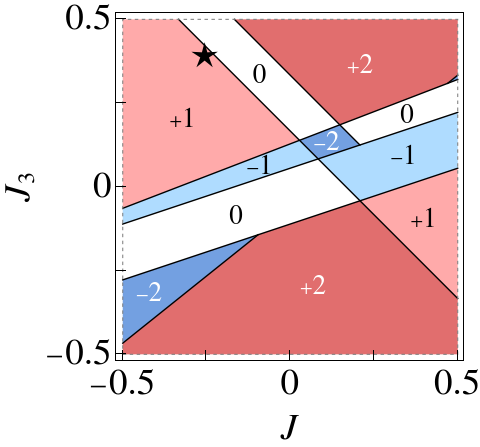}} \qquad \qquad
\subfloat[]{\label{figure:hall3}
\includegraphics[scale=0.3]{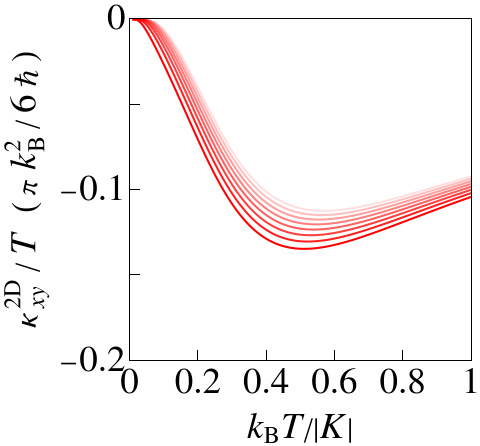}}
\caption{(a) Topological phase diagram of the polarized state under a magnetic field $\mathbf{h} \parallel a$, in the parameter space $K=-1$, $\Gamma = 1$, $\Gamma'=0$, and $J,J_3 \in [-0.5,0.5]$. Dark red, light red, white, light blue, and dark blue areas indicate the Chern number of the lower magnon band $\nu = +2, +1, 0, -1$, and $-2$, respectively, while black lines indicate the vanishing of the band gap. Black star marks the candidate parametrization of $\alpha$-RuCl$_3$ proposed by Ref.~\cite{PhysRevB.93.214431}. (b) Magnon thermal Hall conductivity of the aforementioned parametrization, with $K$ set to -1 and other interactions scaled accordingly, at field strengths from $h=2.02$ to $2.16$ in steps of 0.02. Lighter colors indicate higher fields. $S=1/2$ is used.}
\end{figure}

Here, we specialize to the parameter space defined by $K=-1$, $\Gamma = 1$, $\Gamma'=0$, $ -0.5 \leq J \leq 0.5, -0.5 \leq J_3 \leq 0.5$ and the field angle $\beta = 0$ (i.e., $\mathbf{h} \parallel a$), where analytical progress is possible. Such a choice of parameter space is motivated by the candidate parametrization $(J,K,\Gamma,J_3)=(-1.7,-6.7,6.6,2.7) \, \mathrm{meV}$ of $\alpha$-RuCl$_3$ proposed by Ref.~\cite{PhysRevB.93.214431}, where $\Gamma \approx - K > 0$ and $\Gamma'$ is negligible. Note that $c_3=-1/2$, $c_4=0$, $c_5=5/6$, $c_6=0$, and $c_7=-\sqrt{2/3}$ are fixed by the values of $K$, $\Gamma$, $\Gamma'$, and $\beta$. To construct the topological phase diagram, we first identify the parameters where the band gap closes, and then compute the Chern numbers at the parameters where the band gap is finite, similar to what we have done in Sec.~\ref{supply:lswt}. The result is plotted in Fig.~\ref{figure:parameter3}. The candidate parametrization is located within the $\nu = +1$ regime, which gives rise to a negative and sizable thermal Hall conductivity as expected, see Fig.~\ref{figure:hall3}. We also note that there exist $\nu = \pm 2$ regimes in Fig.~\ref{figure:parameter3}, which do not appear in the $J K \Gamma \Gamma'$ model studied in the main text. \\

Below, we provide the details of determining the critical regions in the aforementioned parameter space. Let $\Delta (\mathbf{k})=0$. Proposition 1 implies that the field dependent and independent parts of \eqref{Dksquareredefine} are both zero, leading to the conditions
\begin{subequations}
\begin{align}
& c_2 f_\mathbf{k} + c_8 t_\mathbf{k} + c_3 = 0 , \label{fielddependentcomplex} \\
\begin{split} \label{fieldindependentcomplex}
&  - 4 \big\{ \mathrm{Im} \big[ c_3 (c_2 f_\mathbf{k} + c_8 t_\mathbf{k} + c_5 f_\mathbf{k}) + c_5 c_8 f_\mathbf{k} t_\mathbf{k}^* \big] \big\}^2 + 32 \, \mathrm{Im} \big[ c_7 g_\mathbf{k}^* (c_2 f_\mathbf{k} + c_8 t_\mathbf{k} + c_5 f_\mathbf{k}) \big] \mathrm{Im} \big[ c_3 c_7 g_\mathbf{k} \big] \\
& + 16 \big\{ \mathrm{Im} \big[ c_5 c_7 f_\mathbf{k} g_\mathbf{k}^* \big] \big\}^2 - 16 \big\{ \mathrm{Im} \big[ c_7 g_\mathbf{k}^* (c_2 f_\mathbf{k} + c_8 t_\mathbf{k}) \big] \big\}^2 = 0 .
\end{split}
\end{align}
\end{subequations}
Substituting \eqref{fielddependentcomplex} in \eqref{fieldindependentcomplex} yields
\begin{equation} \label{fieldindependentintermediate}
16 c_7^2 \big\{ c_3 \mathrm{Im} \big[ g_\mathbf{k}^* \big] - c_5 \mathrm{Im} \big[ f_\mathbf{k} g_\mathbf{k}^* \big] \big\}^2 = 0 \implies (c_3 - c_5) (\sin k_1 - \sin k_2) - 2 c_5 \sin (k_1 - k_2) = 0 .
\end{equation}
Substituting $c_3=-1/2$ and $c_5 = 5/6$ in \eqref{fieldindependentintermediate}, writing down the real and imaginary parts of \eqref{fielddependentcomplex} separately, we obtain the following set of equations,
\begin{subequations}
\begin{align}
& 4 (\sin k_1 - \sin k_2) + 5 \sin (k_1 - k_2) = 0 , \label{fieldindependentsimple} \\
& c_2 (\sin k_1 + \sin k_2) + c_8 \sin (k_1 + k_2) = 0 , \label{fielddependentimag} \\
& c_3 + c_2 ( 1 + \cos k_1 + \cos k_2 ) + c_8 [ 2 \cos (k_1 - k_2) + \cos (k_1 + k_2) ] = 0 . \label{fielddependentreal}
\end{align}
\end{subequations}
We solve \eqref{fieldindependentsimple}-\eqref{fielddependentreal} on a case by case basis. \\

\noindent \quad \textbf{Case 1.}~$\sin k_1 = 0$, or $k_1 = 0, \pi$. Then,  \eqref{fieldindependentsimple} implies $\sin k_2 = 0$, or $k_2 = 0, \pi$. Eq.~\eqref{fielddependentimag} is satisfied by all four possible combinations of $k_1$ and $k_2$.

\noindent \quad \quad \textbf{Case 1.1.}~$(k_1, k_2) = (0,0)$. Eq.~\eqref{fielddependentreal} implies $c_3 + 3 c_2 + 3 c_8 = 0$.

\noindent \quad \quad \textbf{Case 1.2.}~$(k_1, k_2) = (0,\pi)$. Eq.~\eqref{fielddependentreal} implies $c_3 + c_2 - 3 c_8 = 0$.

\noindent \quad \quad \textbf{Case 1.3.}~$(k_1, k_2) = (\pi,0)$. Eq.~\eqref{fielddependentreal} implies $c_3 + c_2 - 3 c_8 = 0$, which is same as Case 1.2.

\noindent \quad \quad \textbf{Case 1.4.}~$(k_1, k_2) = (\pi,\pi)$. Eq.~\eqref{fielddependentreal} implies $c_3 - c_2 + 3 c_8 = 0$.

\noindent \quad \textbf{Case 2.}~$\sin k_1 \neq 0$.

\noindent \quad \quad \textbf{Case 2.1.}~$k_1 = k_2$. Then, \eqref{fieldindependentsimple} is satisfied. Eq.~\eqref{fielddependentimag} becomes $2 c_2 \sin k_1 + 2 c_8 \sin k_1 \cos k_1 = 0$, which implies $c_2 + c_8 \cos k_1 = 0$. If $c_8 = 0$, then $c_2 = 0$, but \eqref{fielddependentreal} cannot be satisfied as $c_3 \neq 0$. Thus $c_8 \neq 0$ and $\cos k_1 = - c_2 / c_8$, which only admits a solution when $\lvert c_2 / c_8 \rvert < 1$. We have excluded $\lvert c_2 / c_8 \rvert = 1$ as it implies $\sin k_1 = 0$, a contradiction. Substituting $\cos k_1 = - c_2 / c_8$ in \eqref{fielddependentreal} yields $c_3 + c_2 + c_8 = 0$.

\noindent \quad \quad \textbf{Case 2.2.}~$k_1 = - k_2$. Then, \eqref{fielddependentimag} is satisfied. Eq.~\eqref{fieldindependentsimple} becomes $8 \sin k_1 + 10 \sin k_1 \cos k_1 = 0$, which implies $\cos k_1 = - 4/5$. Substituting this in \eqref{fielddependentreal} yields $25 c_3 - 15 c_2 + 39 c_8 = 0$.

\noindent \quad \quad \textbf{Case 2.3.}~$k_1 \neq \pm k_2$. Define $k_\pm = k_1 \pm k_2$. Then, $\sin ( k_\pm / 2 ) \neq 0$. Eq.~\eqref{fieldindependentsimple} becomes $8 \cos (k_+ / 2) \sin (k_- / 2) + 10 \sin (k_- / 2) \cos (k_- / 2) = 0$, which implies 
\begin{equation} \label{fieldindependent23}
4 \cos \frac{k_+}{2} + 5 \cos \frac{k_-}{2} = 0 .
\end{equation}
Eq.~\eqref{fielddependentimag} becomes $2 c_2 \sin (k_+ / 2) \cos (k_- / 2) + 2 c_8 \sin (k_+ / 2) \cos (k_+ / 2) = 0$, which implies
\begin{equation} \label{fielddependentimag23}
c_2 \cos \frac{k_-}{2} + c_8 \cos \frac{k_+}{2} = 0 .
\end{equation}
Eqs.~\eqref{fieldindependent23} and \eqref{fielddependentimag23} together imply $( 4 c_2 - 5 c_8 ) \cos (k_+ / 2) = 0$. We now claim that $\cos (k_+ / 2) \neq 0$. Suppose that the contrary is true, i.e., $\cos (k_+ / 2) = 0$, which implies $\cos (k_- / 2) =0$ by \eqref{fieldindependent23}. These yield $k_\pm = n_\pm \pi$ for some odd integers $n_\pm$, which then imply $k_1 = n_1 \pi$ with $n_1 \in \mathbb{Z}$, leading to $\sin k_1 = 0$, a contradiction. Thus $\cos (k_+ / 2) \neq 0$ and $4 c_2 - 5 c_8 = 0$. Substituting \eqref{fieldindependent23} in \eqref{fielddependentreal} yields
\begin{equation} \label{fielddependentreal23}
c_3 + c_2 - 3 c_8 + \Big( - \frac{8}{5} c_2 + \frac{114}{25} c_8 \Big) \cos^2 \frac{k_+}{2} = 0.
\end{equation}
If the coefficient of $\cos^2 (k_+ / 2)$ is zero, then \eqref{fielddependentreal23} reduces to $c_3 + c_2 - 3 c_8 = 0$, a condition that has already been found in Cases 1.2 and 1.3. Therefore, we can assume the coefficient to be nonzero, such that $\cos^2 (k_+ / 2)$ admits a solution when
\begin{equation} \label{inequality23}
0 < \frac{25 ( c_3 + c_2 - 3 c_8 )}{40 c_2 - 114 c_8} < 1 .
\end{equation}
We have excluded $\cos^2 (k_+ / 2) = 0$ and $1$, since the latter implies $\sin (k_+ / 2) = 1$, a contradiction. \\

Collecting all the results, for the specfic parameter space and field angle under consideration, the critical regions satisfy one of the following equations: (I) $c_3 + 3 c_2 + 3 c_8 = 0$, (II) $c_3 + c_2 - 3 c_8 = 0$, (III) $c_3 - c_2 + 3 c_8 = 0$, (IV) $c_3 + c_2 + c_8 = 0$ if $\lvert c_2 / c_8 \rvert < 1$, (V) $25 c_3 - 15 c_2 + 39 c_8 = 0$, and (VI) $4 c_2 - 5 c_8 = 0$ if $40 c_2 - 114 c_8 \neq 0$ and \eqref{inequality23} holds. We can then compute the Chern numbers at parameters away from (I-VI) with the method described in Sec.~\ref{supply:chern}.

\section{Beyond Linear Spin Wave Theory}

Our main results in this work are derived from linear spin wave theory (LSWT), which yields well defined band structures comprising single-magnon states. This allows us to study the magnon band topology for generic model parameters and field angles, and relate it to the thermal Hall conductivity given by the formula \eqref{thermalhallconductivity}, which itself is also derived under the assumption of a non-interacting, quadratic Hamiltonian. When the system is near the phase transition between the high-field polarized state and some other low-field state, quantum fluctuations may be strong such that LSWT breaks down and one has to deal with magnon interactions. While addressing the interacting regime in full is beyond the scope of this work, we state that LSWT is nonetheless a valid approximation at sufficiently high fields where the single-magnon dispersion do not overlap with the multi-magnon continuum \cite{annurev-conmatphys-031620-104715}, in particular the two-magnon continuum. We demonstrate this expectation by treating higher order quantum fluctuations perturbatively via nonlinear spin wave theory (NLSWT), where we perform the Holstein-Primakoff expansion of the Hamiltonian beyond the harmonic level,
\begin{equation}
H = H^{(0)} + H^{(2)} + H^{(3)} + H^{(4)} + \ldots
\end{equation}
Upon factoring out $S^2$, $H^{(0)} \sim 1$ is the classical energy, $H^{(2)} \sim 1/S$ is the linear spin wave Hamiltonian, and $H^{(3)} \sim 1/S^{3/2}$ and $H^{(4)} \sim 1/S^2$ are the cubic and quartic anharmonicities. \\

\begin{figure}
\includegraphics[scale=0.28]{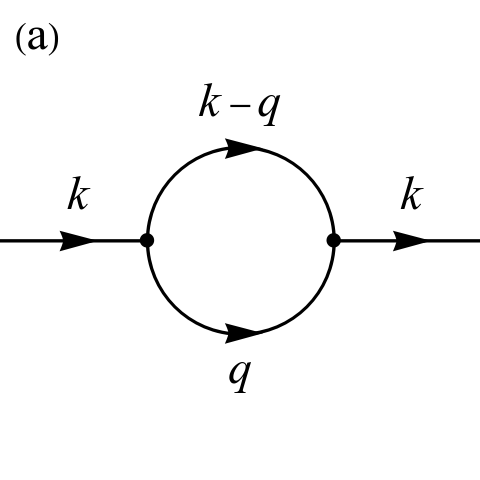} \quad
\includegraphics[scale=0.25]{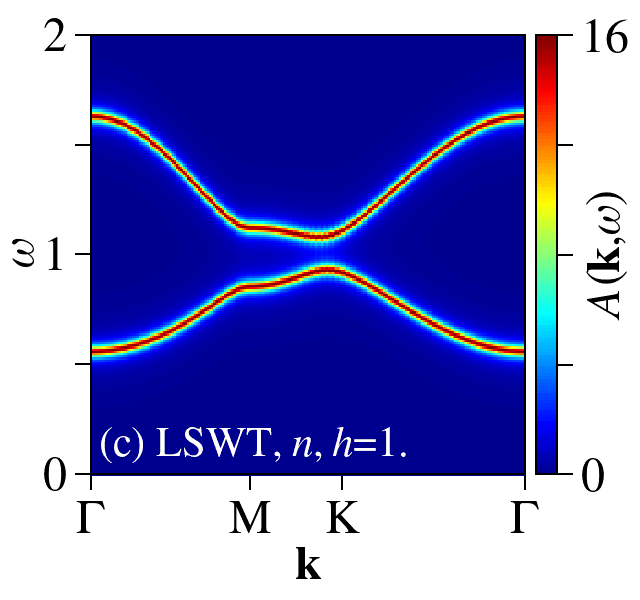} \quad
\includegraphics[scale=0.25]{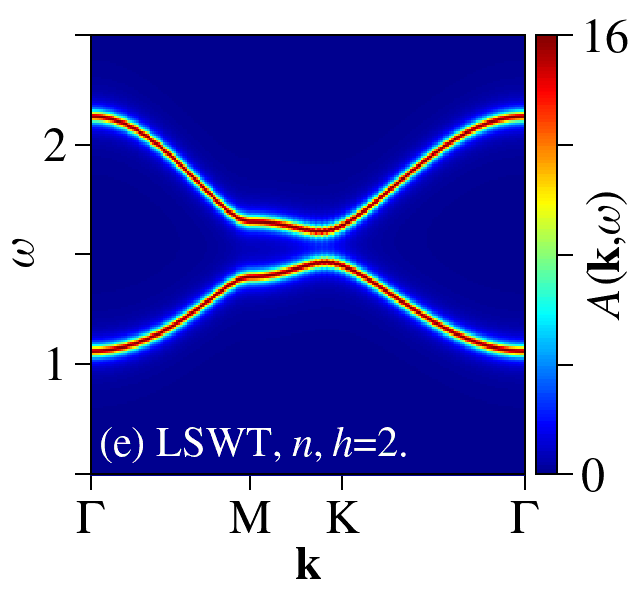} \\
\includegraphics[scale=0.28]{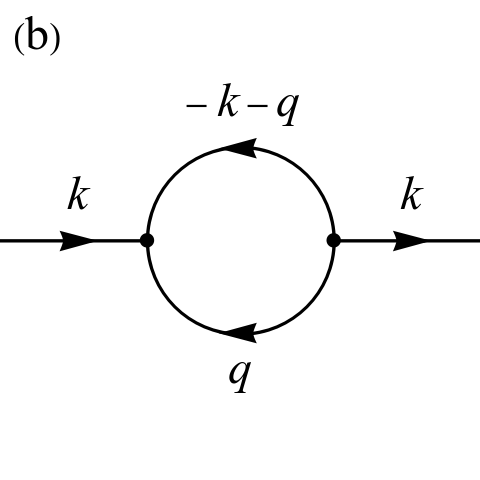} \quad
\includegraphics[scale=0.25]{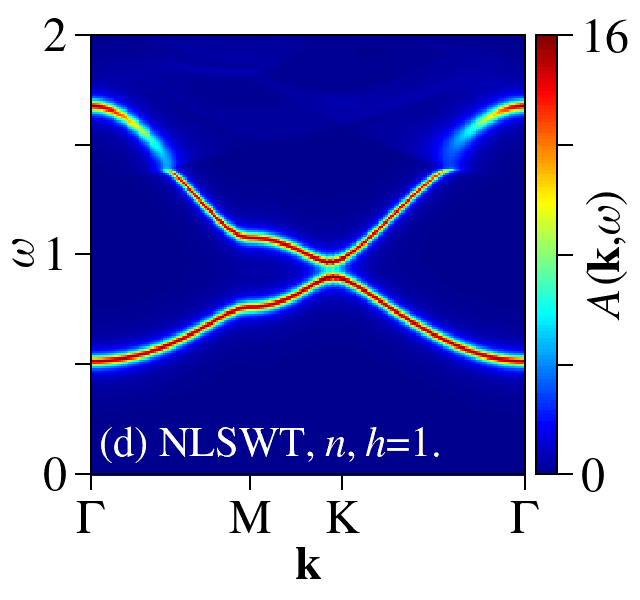} \quad
\includegraphics[scale=0.25]{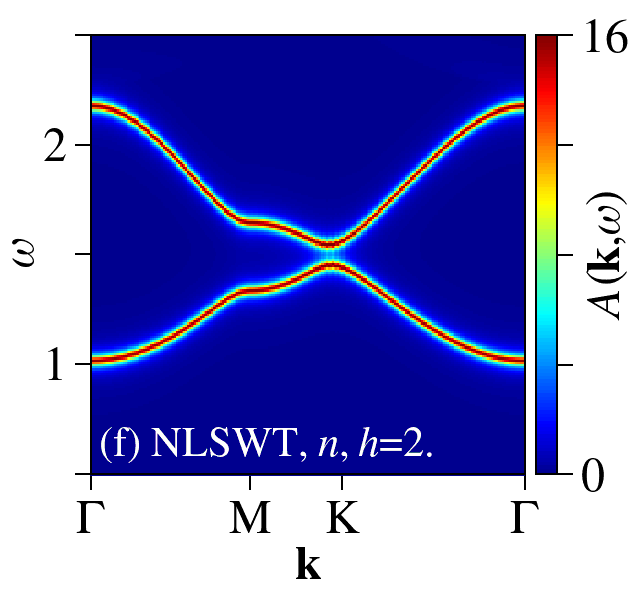}
\caption{\label{figure:spectral}Lowest order contributions to the self-energy from the (a) decay and (b) source terms in the three-magnon Hamiltonian, see \eqref{cubicanharmony}, \eqref{decayselfenergy}, and \eqref{sourceselfenergy}. Spectral functions calculated from linear spin wave theory (LSWT) for the parametrization $n$ under a magnetic field $\mathbf{h} \parallel a$, at (c) $h=1$ and (e) $h=2$. Spectral functions calculated from nonlinear spin wave theory (NLSWT) for the parametrization $n$ under a magnetic field $\mathbf{h} \parallel a$, at (d) $h=1$ and (f) $h=2$. $S=1/2$ is used.}
\end{figure}

\begin{figure}
\includegraphics[scale=0.25]{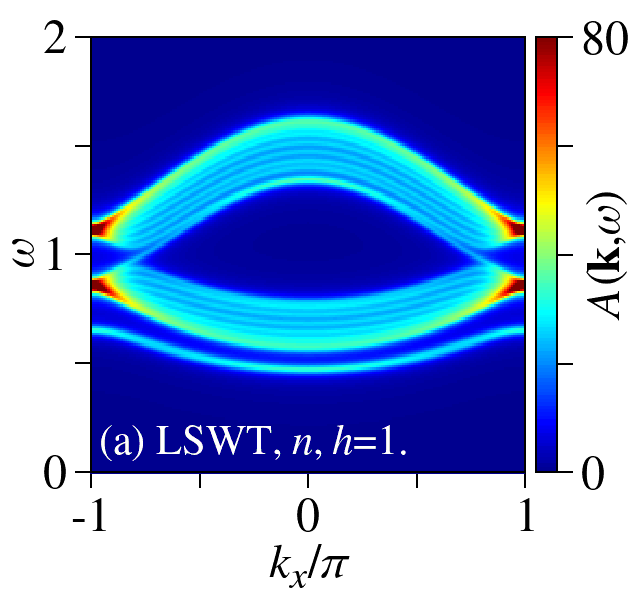}
\includegraphics[scale=0.25]{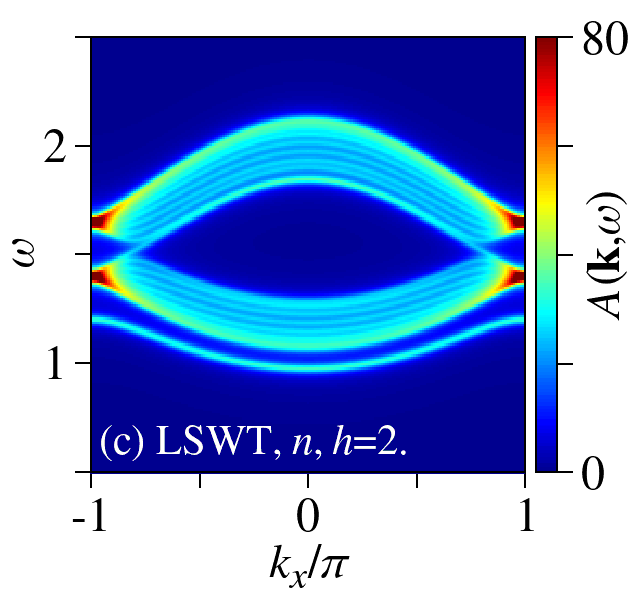}
\includegraphics[scale=0.25]{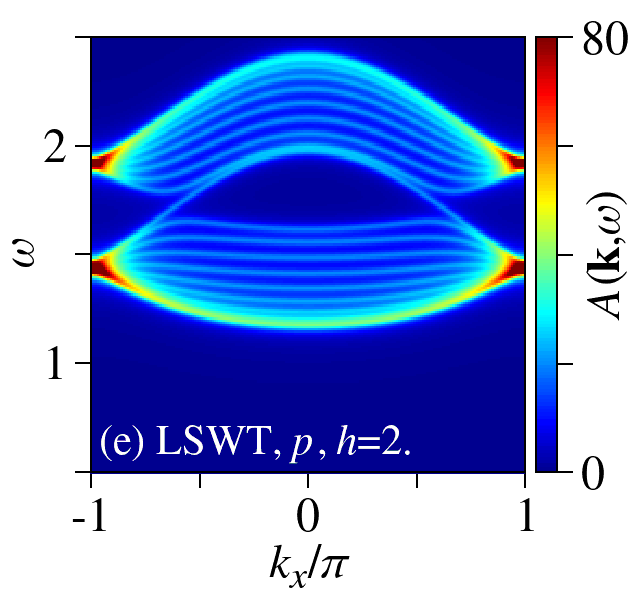}
\includegraphics[scale=0.25]{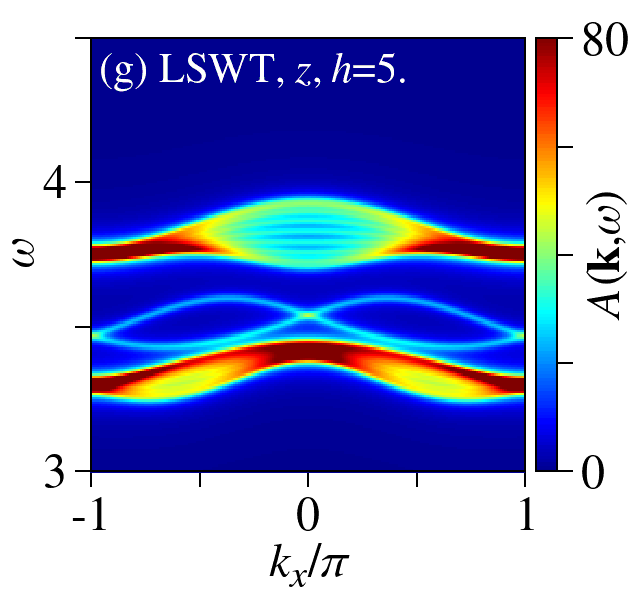} \\
\includegraphics[scale=0.25]{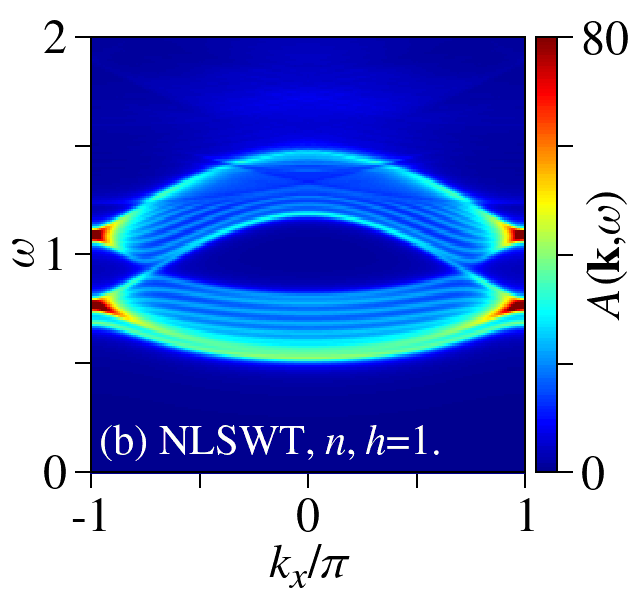}
\includegraphics[scale=0.25]{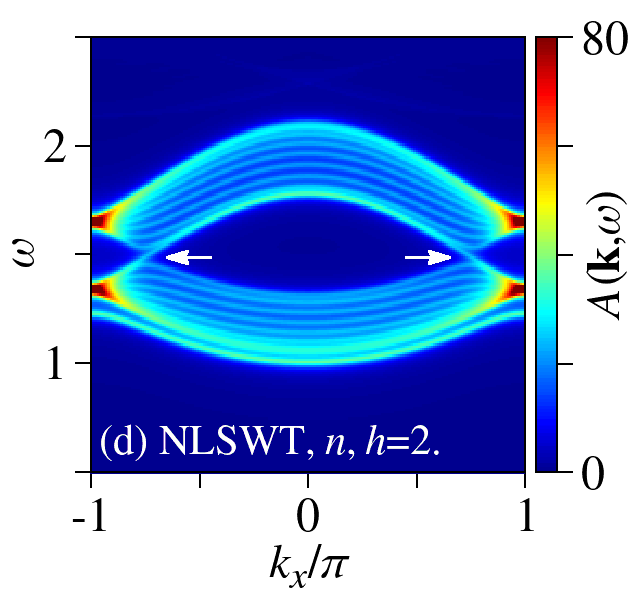}
\includegraphics[scale=0.25]{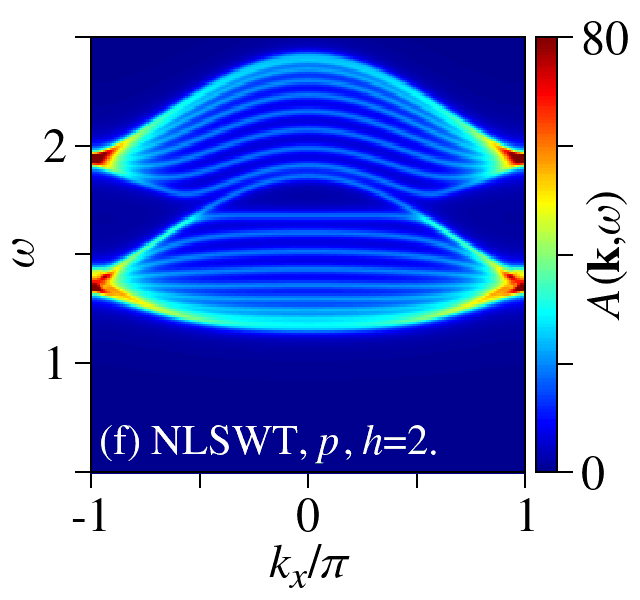}
\includegraphics[scale=0.25]{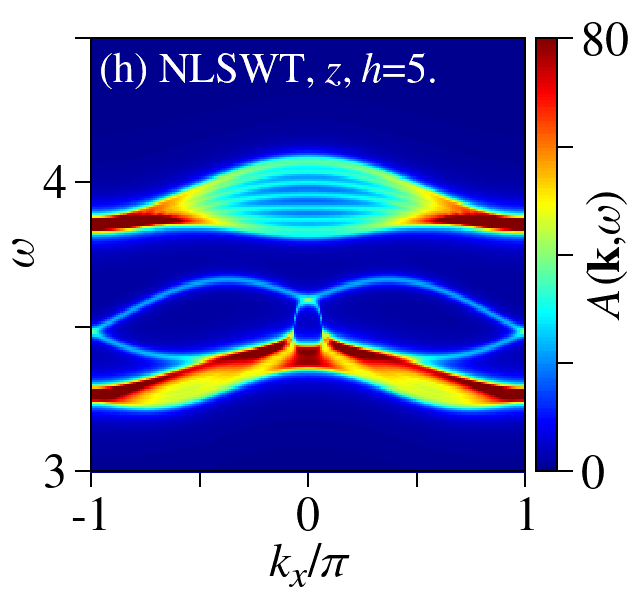}
\caption{\label{figure:slab}Spectral functions on a slab geometry calculated from linear spin wave theory (LSWT) for various parametrizations under a magnetic field $\mathbf{h} \parallel a$: (a) $n$ at $h=1$, (c) $n$ at $h=2$, (e) $p$ at $h=2$, and (g) $z$ at $h=5$. Spectral functions on a slab geometry calculated from nonlinear spin wave theory (NLSWT) for various parametrizations under a magnetic field $\mathbf{h} \parallel a$: (b) $n$ at $h=1$, (d) $n$ at $h=2$, (f) $p$ at $h=2$, and (h) $z$ at $h=5$. $S=1/2$ is used. In (d), we explicitly indicate the magnon modes that connect the upper and lower bulk bands with white arrows.}
\end{figure}

A useful quantity to study how the magnon spectrum is affected by interactions is the spectral function $A (\mathbf{k} , \omega)$, which is proportional to the imaginary part of the Green's function $G (\mathbf{k} , \omega)$. We calculate self-energy corrections to $G (\mathbf{k} , \omega)$ up to $\mathcal{O} (1/S^2)$, and plot $A (\mathbf{k} , \omega)$ for selected parametrizations under a magnetic field along the $a$ axis, with the coupling constants scaled such that $K=-1$ as in Fig.~\ref{figure:parameter} of the main text. For the parametrization $n$ at $h=1$, part of the upper band is severed from the rest and broadened by the two-magnon continuum, see Fig.~\ref{figure:spectral}d. Increasing the field to $h=2$, we recover well defined single-magnon energies, though they are slightly renormalized, see Fig.~\ref{figure:spectral}f. The corresponding results from LSWT are also presented for comparison, see Figs.~\ref{figure:spectral}c and \ref{figure:spectral}e. More remarkably, the connectivity of the upper and lower bulk bands in a slab geometry as calculated from NLSWT (lower panel of Fig.~\ref{figure:slab}) is consistent with the presence or absence of chiral edge modes, which signify nontrivial or trivial band topology, as calculated from LSWT (upper panel of Fig.~\ref{figure:slab}), at least in the high field regime. For the parametrizations $n$ and $p$ where LSWT yields topological magnons, there exist energy states that run continuously from the lower to upper bulk bands in NLSWT, see Figs.~\ref{figure:slab}b, \ref{figure:slab}d, and \ref{figure:slab}f. For the parametrization $z$ where LSWT yields trivial magnons, the upper and lower bulk bands are disconnected from each other in NLSWT, see Fig.~\ref{figure:slab}h. Note that despite the partial overlap between the upper bulk bands and the two magnon continuum at high energies in Fig.~\ref{figure:slab}b, the connection survives. \\

Below, we sketch the NLSWT calculations, noting that detailed and extensive accounts can be found in the existing literature, see Refs.~\cite{PhysRevB.79.144416,RevModPhys.85.219,PhysRevB.88.094407,PhysRevB.106.214411,PhysRevB.109.014424} for example. Our goal is to evaluate the diagonal part of the (retarded) Green's function at zero temperature. Including self-energy corrections up to $\mathcal{O}(1/S^2)$, we have \cite{PhysRevB.88.094407,PhysRevB.106.214411}
\begin{equation} \label{greenfunction}
G_n (\mathbf{k} , \omega) = \frac{1}{\omega - \varepsilon_{n \mathbf{k}} - \Sigma_n (\mathbf{k}, \omega) + i0^+} ,
\end{equation}
where $n = 1,2$ is the band index (not to be confused with the candidate parametrization), $\varepsilon_{n \mathbf{k}}$ is the single-magnon dispersion from LSWT [which was previously denoted as $\omega_\pm (\mathbf{k})$], and $\Sigma_n (\mathbf{k}, \omega)$ is the diagonal part of the self-energy. The corresponding spectral function is given by
\begin{equation} \label{spectralfunction}
A_n (\mathbf{k}, \omega) = - \frac{1}{\pi} \, \mathrm{Im} \, G_n (\mathbf{k}, \omega)
\end{equation}
and we define $A (\mathbf{k} , \omega) = \sum_n A_n (\mathbf{k} , \omega)$. We remark that while our discussion here assumes a genuine two-dimensional system with periodic boundary conditions in both directions, it can be straightforwardly extended to the slab geometry with open boundary condition in one direction and periodic boundary condition in the other. \\

The three-magnon Hamiltonian has the general form \cite{PhysRevB.79.144416,PhysRevB.93.014418,PhysRevB.106.214411,PhysRevB.109.014424}
\begin{equation} \label{cubicanharmony}
H^{(3)} = \frac{1}{2! \sqrt{N}} \sum_{\mathbf{k} \mathbf{q}} \sum_{l m n} \big( \Phi^{lmn}_{\mathbf{q}, \mathbf{k}-\mathbf{q}; \mathbf{k}} \gamma_{l \mathbf{q}}^\dagger \gamma_{m \mathbf{k}-\mathbf{q}}^\dagger \gamma_{n \mathbf{k}} + \mathrm{h.c.} \big) + \frac{1}{3! \sqrt{N}} \sum_{\mathbf{k} \mathbf{q}} \sum_{lmn} \big( \Xi^{lmn}_{\mathbf{q}, -\mathbf{k}-\mathbf{q}, \mathbf{k}} \gamma_{l \mathbf{q}}^\dagger \gamma_{m -\mathbf{k}-\mathbf{q}}^\dagger \gamma_{n \mathbf{k}}^\dagger + \mathrm{h.c.} \big) ,
\end{equation}
where $\gamma$ is the Bogoliubov quasiparticle defined via $\Psi_\mathbf{k} = T_\mathbf{k} \Gamma_\mathbf{k}$ and $\Gamma_\mathbf{k} = (\gamma_{1 \mathbf{k}}, \gamma_{2 \mathbf{k}}, \gamma_{1 - \mathbf{k}}^\dagger, \gamma_{2 - \mathbf{k}}^\dagger)$, see the beginning of Sec.~\ref{supply:lswt}, and the symmetrized vertices $\Phi^{lmn}_{\mathbf{q}, \mathbf{k}-\mathbf{q}; \mathbf{k}}$ and $\Xi^{lmn}_{\mathbf{q}, -\mathbf{k}-\mathbf{q}, \mathbf{k}}$ depend on the coupling constants and the Bogoliubov transformation. The first and second terms in \eqref{cubicanharmony} are known as the decay and source terms, respectively. Their lowest order contributions to the diagonal part of the self-energy are \cite{PhysRevB.79.144416,PhysRevB.88.094407,PhysRevB.106.214411,PhysRevB.109.014424}
\begin{subequations}
\begin{align}
\Sigma_n^{(\mathrm{d})} (\mathbf{k} , \omega) &= \frac{1}{2 N} \sum_{\mathbf{q}} \sum_{lm} \frac{\lvert \Phi^{lmn}_{\mathbf{q}, \mathbf{k}-\mathbf{q}; \mathbf{k}} \rvert^2}{\omega - \varepsilon_{l \mathbf{q}} - \varepsilon_{m \mathbf{k}-\mathbf{q}} + i 0^+} , \label{decayselfenergy} \\
\Sigma_n^{(\mathrm{s})} (\mathbf{k} , \omega) &= - \frac{1}{2 N} \sum_{\mathbf{q}} \sum_{lm} \frac{\lvert \Xi^{lmn}_{\mathbf{q}, -\mathbf{k}-\mathbf{q}, \mathbf{k}} \rvert^2}{\omega + \varepsilon_{l \mathbf{q}} + \varepsilon_{m -\mathbf{k}-\mathbf{q}} - i 0^+} , \label{sourceselfenergy}
\end{align}
\end{subequations}
which are diagrammatically represented as Figs.~\ref{figure:spectral}a and \ref{figure:spectral}b. On the other hand, the contribution of the four-magnon Hamiltonian to the self-energy at $\mathcal{O} (1/S^2)$ is equivalent to a Hartree-Fock approximation, where one performs a mean field decoupling of $H^{(4)}$ and obtains a quadratic Hamiltonian $\delta H^{(2)}$ \cite{PhysRevB.79.144416,PhysRevB.102.155134}. We further implement a self-consistent scheme \cite{PhysRevB.98.060404,PhysRevB.106.094431,PhysRevB.109.014424} as follows. Neglecting $H^{(3)}$ at first, we diagonalize $H^{(2)} + \delta H^{(2)}$ by a Bogoliubov transformation, which is then used to evaluate the Hartree-Fock averages. This leads to a modified $\delta H^{(2)}$, and we iterate this procedure until convergence is reached. Such a self-consistent treatment actually goes beyond $\mathcal{O} (1/S^2)$, but it has the advantage of regularizing unphysical divergences that may appear in the magnon spectrum at the strict $1/S^2$ order \cite{PhysRevB.109.014424}. The renormalized single-magnon dispersion $\overline{\varepsilon}_{n \mathbf{k}}$ and Bogoliubov transformation $\overline{T}_\mathbf{k}$ obtained from diagonalizing $H^{(2)} + \delta H^{(2)}$ upon convergence are then used to evaluate the self-energy contributions \eqref{decayselfenergy} and \eqref{sourceselfenergy} from $H^{(3)}$. Finally, we replace $\varepsilon_{n \mathbf{k}}$ by $\overline{\varepsilon}_{n \mathbf{k}}$ and let $\Sigma_n = \Sigma_n^{(\mathrm{d})} + \Sigma_n^{(\mathrm{s})}$ in the Green's function \eqref{greenfunction}. In the numerical calculations, we use $5 \times 10^{-3}$ for $0^+$ in \eqref{decayselfenergy} and \eqref{sourceselfenergy}, and $2 \times 10^{-2}$ for $0^+$ in \eqref{greenfunction}. \\

We also provide the details of the slab geometry used in the calculations of Figs.~\ref{figure:slab}a-\ref{figure:slab}h. We choose the primitive lattice vectors such that $\mathbf{a}_1$ is aligned with the $a$ axis, along which the magnetic field is applied, and $\mathbf{a}_2$ is rotated by $\pi/3$ from the $a$ axis. The slab geometry is essentially infinite along the $\mathbf{a}_1$ direction, which defines the momentum $k_x$, and finite along the $\mathbf{a}_2$ direction with $10$ unit cells and open boundary condition.

\end{document}